\documentclass[graybox,oneside,openany]{svmult}

\usepackage[utf8]{inputenc}
\usepackage{type1cm}        
\usepackage{makeidx}         
\usepackage{graphicx}        
\usepackage{titlesec}
\usepackage{multicol}        
\usepackage[bottom]{footmisc}
\usepackage[backref=page]{hyperref}
\usepackage{amssymb}
\usepackage{bm}
\usepackage{newtxtext}       %
\usepackage{newtxmath}       
\usepackage{todonotes}

\graphicspath{{./figures/}}

\makeindex             

\begin{document}

\title*{Chapter 58: How does artificial intelligence contribute to iEEG research?}
\author{Julia Berezutskaya, Anne-Lise Saive, Karim Jerbi \&  Marcel van Gerven}
\institute{Julia Berezutskaya 
    \at Donders Institute for Brain, Behavior \& Cognition, 
            Radboud University, Thomas van Aquinostraat 4, 6525 GD Nijmegen, the Netherlands
    \email{yuliya.berezutskaya@donders.ru.nl}
\and Anne-Lise Saive 
    \at Institut Paul Bocuse Research Center, Lyon, France, 
    \email{anne-lise.saive@institutpaulbocuse.com}
\and Karim Jerbi 
    \at Cognitive \& Computational Neuroscience Lab, 
        University of Montreal, Pavillon Marie-Victorin 90, avenue Vincent d'Indy, Quebec, Canada, 
    \email{karim.jerbi@umontreal.ca}
\and Marcel van Gerven 
    \at Donders Institute for Brain, Behavior \& Cognition, 
            Radboud University, Thomas van Aquinostraat 4, 6525 GD Nijmegen, the Netherlands 
    \email{marcel.vangerven@donders.ru.nl}
}
%
%

{\def\addcontentsline#1#2#3{}\maketitle{\vspace{-3em}}}

\begin{center}
\paragraph{\emph{Note:} This chapter is forthcoming in \emph{Intracranial EEG for Cognitive Neuroscience}}
\end{center}
\abstract{Artificial intelligence (AI) is a fast-growing field focused on modeling and machine implementation of various cognitive functions with an increasing number of applications in computer vision, text processing, robotics, neurotechnology, bio-inspired computing and others. In this chapter, we describe how AI methods can be applied in the context of intracranial electroencephalography (iEEG) research. IEEG data is unique as it provides extremely high-quality signals recorded directly from brain tissue. Applying advanced AI models to these data carries the potential to further our understanding of many fundamental questions in neuroscience. At the same time, as an invasive technique, iEEG lends itself well to long-term, mobile brain-computer interface applications, particularly for communication in severely paralyzed individuals. We provide a detailed overview of these two research directions in the application of AI techniques to iEEG. That is, (1) the development of computational models that target fundamental questions about the neurobiological nature of cognition (AI-iEEG for neuroscience) and (2) applied research on monitoring and identification of event-driven brain states for the development of clinical brain-computer interface systems (AI-iEEG for neurotechnology). We explain key machine learning concepts, specifics of processing and modeling iEEG data and details of state-of-the-art iEEG-based neurotechnology and brain-computer interfaces.}


\setcounter{tocdepth}{3}
\setcounter{secnumdepth}{4}
\titleformat{\paragraph}
{\normalfont\normalsize\bfseries}{\theparagraph}{1em}{}
\titlespacing*{\paragraph}
{0pt}{3.25ex plus 1ex minus .2ex}{1.5ex plus .2ex}
\tableofcontents
\newpage



\section{AI-iEEG for neuroscience} 
\label{sec:science}

The field of computational cognitive neuroscience uses mathematical models to describe neural processes underlying cognition and behaviour~\cite{Ashby2011a,OReilly,VanGerven2017b}. Advances in machine learning and the rapid increase in computational power have made it possible to apply sophisticated analysis methods to large amounts of brain data collected via increasingly sophisticated recording techniques~\cite{Churchland2016}. 

Inspired by the computational metaphor of the brain as an information processing device, this has led to the emergence of so-called \emph{encoding models} and \emph{decoding models} in cognition and perception~\cite{holdgraf2017encoding, van2017primer, kriegeskorte2019interpreting, king2020encoding}. Neural encoding models capture how information is represented and processed in the brain.  
An approach related to encoding models, that has become increasingly popular in cognitive neuroscience, is \emph{representational similarity analysis} (RSA)~\cite{diedrichsen2017representational}. RSA more directly compares stimulus features as encoded in computational models and encoded in patterns of brain activity.
Decoding models also aim to understand the representation of information in the brain but constitute the reverse approach of inferring information features from observed brain activity. Apart from the practical value of decoding approaches for neurotechnology (see Section \ref{sec:technology}), decoding models can also be informative in relating stimulus features to neural signals. 

Given that many cognitive functions, including perception, memory, language and complex sensorimotor behaviour, are supported by large-scale distributed processes in the brain, data from neuroimaging (fMRI) and whole-brain electrophysiological (EEG/MEG) experiments in healthy individuals has been the primary source for construction and validation of computational models. The FMRI, MEG and EEG communities have provided large amounts of experimental data used for modeling cognitive processes in the brain~\cite{hanke2014high, wakeman2015multi,  seeliger2019large, schoffelen2019204,  nastase2021narratives,Armeni2022}. 

However, limitations of non-invasive brain recording modalities have created a need for the use of intracranial data. This is because iEEG offers a number of benefits compared to non-invasive modalities, such as sampling directly from the cortical tissue, high signal-to-noise ratio and high temporal and spatial resolution. Recent work has seen successful efforts in creation and validation of encoding and decoding models, RSA and unsupervised models of neural dynamics using iEEG.

In this section we will describe encoding and decoding models, and include discussion of RSA and complex temporal dynamics in iEEG data. First, we will discuss linear encoding models for iEEG signals, including hand-engineered and deep-learning-based feature extraction for these models, and emergence of advanced non-linear neural encoding models. Then, we will examine RSA applications to iEEG. We will finish the section with a description of decoding models applied to iEEG data, including multivariate pattern analysis, and a discussion of associated temporal generalization concerns. 

\subsection{Encoding models of perception and cognition} 

\subsubsection{Linear encoding models}

A linear neural encoding model represents a linear mapping of a set of \emph{features} (model input) to the observed brain activity (model output, see also Chapter 4). Fitting such a model is based on the assumption that the brain activity is a weighted linear combination of features and normally distributed random noise:
\begin{equation}
\mathbf{{Y}}=\mathbf{X}\mathbf{B}+\bm{\epsilon}\,,\tag{1.1}\label{eq:1.1}
\end{equation}
where $\mathbf{Y} \in \mathbb{R}^{n\times m}$ is a matrix of observed neural responses with $n$ corresponding to the number of observations over time or trials, $m$ corresponding, for example, to the number of iEEG electrodes or fMRI voxels,  $\mathbf{X} \in \mathbb{R}^{n \times p}$ is a matrix of input features over $n$ time points and $p$ feature dimensions, $\mathbf{B}\in \mathbb{R}^{p\times m}$ is a matrix of regression coefficients and $\bm{\epsilon} \sim \mathcal{N}(\mathbf{0}, \bm{\sigma}^2 \mathbf{I})$ is a noise term with $\mathbf{I}$ being the identity matrix.

\begin{figure}[!htbp]
\centering
\includegraphics[width=1\linewidth]{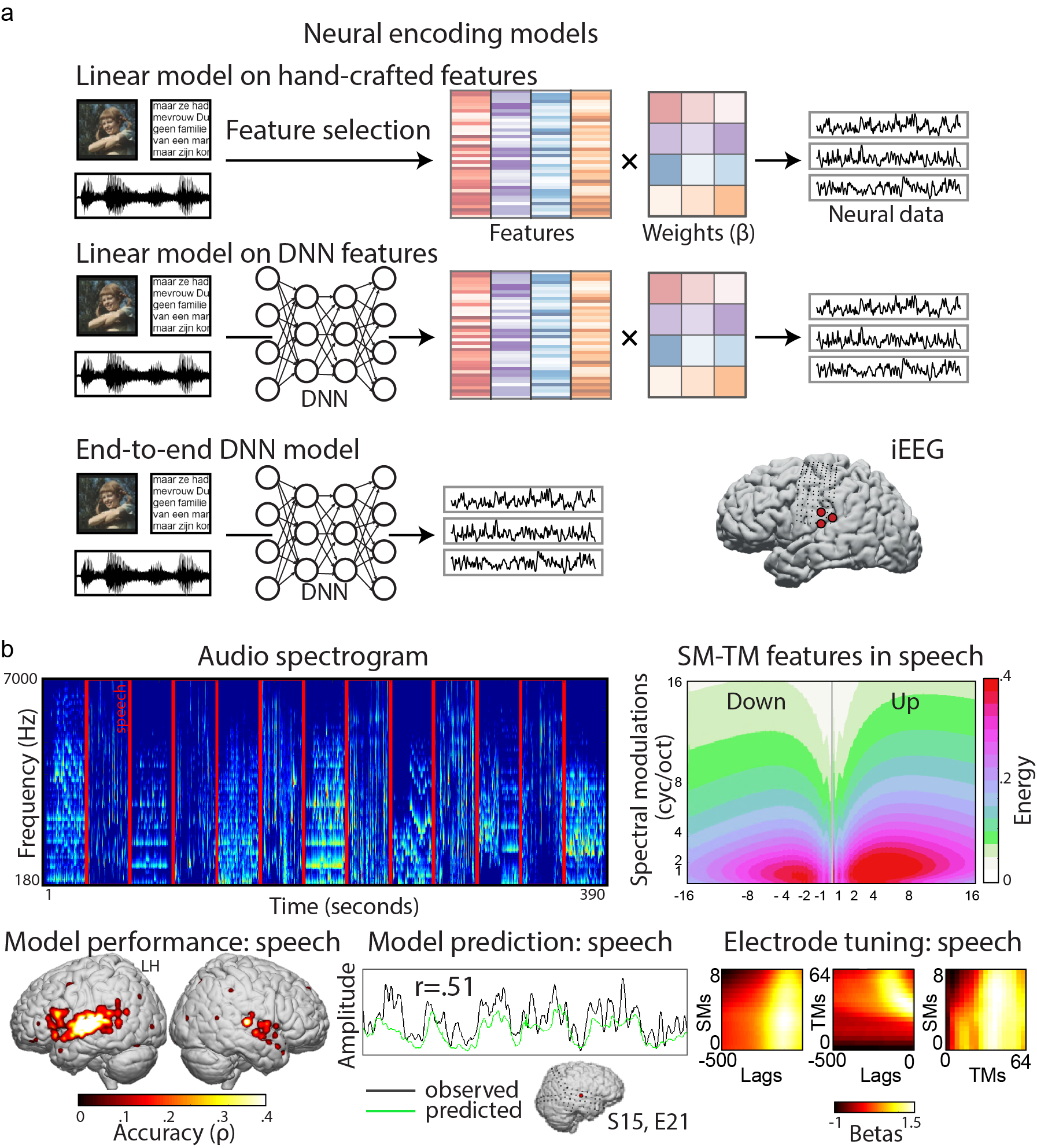}
\caption{\textbf{a}. Overview of neural encoding models: linear encoding model on hand-crafted stimulus features, linear encoding model on features automatically extracted from the stimulus using deep learning models and end-to-end non-linear encoding model based on a deep learning approach. Each model in this example aims to predict high frequency band iEEG responses of three electrodes. \textbf{b}. A linear encoding model that predicts ECoG high frequency band activity during speech dialogues in a short audiovisual film ~\cite{berezutskaya2017neural}. A set of spectrotemporal modulations (SM-TM) was extracted from the sound spectrogram and used to model associated iEEG responses throughout the perisylvian cortex. Each iEEG electrode exhibited tuning to specific features at different time latencies. Adapted from~\cite{berezutskaya2017neural}. Copyright 2017 Berezutskaya et al.}
\label{figure1}
\end{figure}

In this formulation, the model input -- the feature matrix $\mathbf{X}$ -- represents properties of the stimulus that may have triggered the observed brain responses. Linear encoding models are widely used in fitting brain activity evoked by external stimuli (Fig.~\ref{figure1}). IEEG examples that use this simple form of the linear model include neural encoding of visual~\cite{harvey2013frequency, de2020intracranial} and auditory~\cite{mesgarani2014phonetic, hullett2016human, berezutskaya2017neural, oganian2019speech} perception, motor information~\cite{anderson2012electrocorticographic, chartier2018encoding, merrick2022left}, memory, language~\cite{berezutskaya2020high, schrimpf2021neural, goldstein2022shared} and decision making tasks~\cite{saez2018encoding}. 

Typically, models in these studies are fitted to predict the temporal envelope of the high-frequency broadband component ($>$60 Hz) in iEEG data (obtained with time-frequency analysis), as it is known to reflect local neural firing rates evoked by the stimulus~\cite{crone2001induced, ray2008neural}. Alternative encoding approaches have been employed to model frequency-domain representations of iEEG signals~
\cite{miller2009power, winawer2013asynchronous, hermes2015stimulus, gaglianese2017separate}. Time-domain iEEG responses on average show less straightforward modulation by a cognitive task and are more varied compared to frequency-domain responses~\cite{miller2016spontaneous}, and are therefore less typically used in linear fitting with stimulus features.

When working with a time-frequency representation of iEEG signals,  kernel functions $g(\cdot)$ can be used to transform input stimulus features into a new representation that could better reflect temporal iEEG structure, yielding an encoding of the form
\begin{equation}
\mathbf{Y}=g(\mathbf{X})\mathbf{B}+\bm{\epsilon}\,.\tag{1.2}\label{eq:1.2}
\end{equation}
This is similar to fMRI research, where an input feature matrix is convolved with the hemodynamic response function for the general linear model fit. For iEEG, Gaussian kernels can be applied to stimulus features to better approximate smoother iEEG time courses. 
Examples of studies that use such feature transformations include~\cite{salari2018spatial, branco2019high}.

Another property of the spectrotemporal iEEG signal is its high temporal resolution. It is usually addressed in models of continuous iEEG dynamics (as opposed to fitting trial-based data) by inclusion of lags and temporal integration over input features, such that
\begin{equation}
\mathbf{y}_t= 
\bar{\mathbf{B}} \bar{\mathbf{x}}_t +\bm{\epsilon}\,,\tag{1.3}\label{eq:1.3}
\end{equation}
where $\mathbf{y_t} \in \mathbb{R}^{m}$ is a vector of iEEG responses per time point $t$, $\bar{\mathbf{B}} \in \mathbb{R}^{m \times (p \times \tau)}$ a matrix of regression coefficients, $\tau$ is the time lag relative to $t$ 
and $\bar{\mathbf{x}}_t$ a stack of vectors $\mathbf{x}_{t-s}$ with $s \in [0, \tau]$.
Features at multiple lags have been used in models of encoding auditory, speech and movement information in iEEG~\cite{berezutskaya2017neural,chartier2018encoding}.

A standard linear regression problem is typically solved using ordinary least squares estimation of regression coefficients $\mathbf{B}$, which can be achieved analytically: 
\begin{equation}
\mathbf{B} = (\mathbf{X}^\top\mathbf{X})^{-1}\mathbf{X}^\top\mathbf{Y}\,.\tag{1.4}\label{eq:1.4}
\end{equation}
To assess model performance, various metrics are used for comparing predicted and observed brain responses. A popular choice is a Pearson correlation coefficient, which measures the normalized covariance of predictions and their targets. To assess generalization potential of the modeling results, encoding models are trained on a subset of all data (typically 80$-$90\%), and tested on a held-out test set (remaining data). To account for potential autocorrelation in iEEG signals, data from separate recording sessions and different experiments can be used as an independent test set~\cite{berezutskaya2020brain}. Model performance is often \emph{cross-validated} to obtain a more unbiased estimate of encoding accuracy. 

To prevent \emph{overfitting} to training data (and subsequent lack of generalization on test data), \emph{regularization} methods are often used. They provide constraints, or priors, on model coefficients~\cite{diedrichsen2017representational}, resulting in a less flexible model fit and therefore are less prone to overfitting on the training data. Regularization is effective in models that use large numbers of correlated features. A popular method is L2 regularization that constrains $\mathbf{B}$ to small values. A linear model that uses L2 regularization is called a ridge regression and has the following analytical solution:
\begin{equation}
\mathbf{B} = (\mathbf{X}^\top\mathbf{X} + \lambda
\mathbf{I})^{-1}\mathbf{X}^\top\mathbf{Y}\,,\tag{1.5}\label{eq:1.5}
\end{equation}
where $\lambda \geq 0$ is a regularization parameter, typically chosen via nested cross-validation. 

Neural information processing includes high non-linear transformations of sensory input. Therefore, the linear models we often use have severe limitations. They can be mitigated by fitting linear encoding models on non-linear feature representations. The latter can be either manually defined or extracted from computational models that mimic neural processing such as deep and recurrent neural networks.

\subsubsection{Feature engineering for linear encoding models}
\label{subsec:feature_eng}

To fit a linear model of brain activity as a response to external stimuli, one requires a set of features that represent stimulus information. Traditionally and similar to other recording modalities, such as fMRI, MEG and EEG, \emph{hand-engineered features} have been used for this. For example, spectrotemporal features were used in encoding of auditory and speech information in iEEG~\cite{mesgarani2014phonetic, hullett2016human, berezutskaya2017neural}, two-dimensional Gaussian features were used in population receptive field work in vision~\cite{winawer2013asynchronous, harvey2013frequency} and kinematic trajectories and grasp features were used for modeling sensorimotor iEEG signals~\cite{chartier2018encoding, branco2019high}. 

The use of hand-engineered features to represent information means that the choice of features is at the discretion of the researcher. Features could be manually assigned data labels (for example, language labels, such as words or phonemes, classes of visually presented images, broadly defined experimental conditions) or a representation of the stimulus extracted via a preprocessing step or a mathematical model (for example, spectrotemporal audio features or edge-filtered image features). In the case of a sufficiently good fit compared against a model with random predictors, researchers analyse neural tuning profiles, or stimulus receptive fields. Many studies compare the fit of models that use different feature sets in an attempt to identify features that best explain observed brain data~\cite{leeds2013comparing, khaligh2014explaining, wehbe2014simultaneously, yamins2014performance, jozwik2016visual}. 

Despite its overall success, previous work (also with other neural recording techniques) highlighted some problems with using hand-engineered features, specifically, the inability to determine which feature sets should be deemed the optimal representation of information in a specific brain region; and how existing features need to be modified to explain brain responses better. Moreover, the use of hand-engineered features lacked the ability to explain how lower-level feature sets (such as image edges or orientations and spectrogram sound features) transformed to higher-level features (such as object categories and different words) as a result of information processing throughout the brain. Recent progress in \emph{deep learning}, has opened up new possibilities in addressing the problem of feature engineering by minimizing manual feature selection and delegating feature extraction to powerful non-linear models.

In a vanilla deep artificial neural network (DNN), input features are passed to the first layer, where their weighted linear combination goes through a non-linear transformation, called an activation function, to obtain layer activations $\mathbf{a}$. 
This computation is repeated at the next layers. The final layer activations are used to predict the targets. 
 This yields the following sequence of non-linear transformations:
\begin{align}
\mathbf{a}^{(1)}&=g\left({\mathbf{B}^{(1)}}\mathbf{x}_t + \mathbf{b}^{(1)}\right)\tag{1.6}\label{eq:1.6}\\
\mathbf{a}^{(2)}&=g\left({\mathbf{B}^{(2)}}\mathbf{a}^{(1)} + \mathbf{b}^{(2)}\right)\tag{1.7}\label{eq:1.7}\\
\mathbf{\hat{y}}&=g\left({\mathbf{B}^{(k)}}\mathbf{a}^{(k-1)} + \mathbf{b}^{(k)}\right)\tag{1.8}\label{eq:1.8}
\end{align}
where $k \in \mathbb{N}$ is the number of layers and $\mathbf{b}$ is a layer-specific bias vector.

DNNs are trained to optimize an \emph{objective function} associated with the task they need to solve, for example labeling visual objects in the presented image, or predicting next words in a sentence. Minimizing mean squared error is a popular objective function for regression tasks (such as time series prediction), and cross-entropy loss is used in most classification tasks (such as object recognition). To optimize the objective function a \emph{learning algorithm}, typically gradient descent, is applied iteratively. This yields incremental updates of learnable parameters of the DNN, such as weights of its artificial neurons. In most cases, backpropagation is used for efficient gradient calculation. As a result of this learning, DNN layer activations can capture increasingly complex non-linear transformations of simple inputs (such as images, texts and sounds) necessary for solving the task. For a more detailed introduction to DNNs see~\cite{kriegeskorte2015deep, lecun2015deep, kriegeskorte2019neural}. 

\begin{figure}[!htbp]
\centering
\includegraphics[width=1\linewidth]{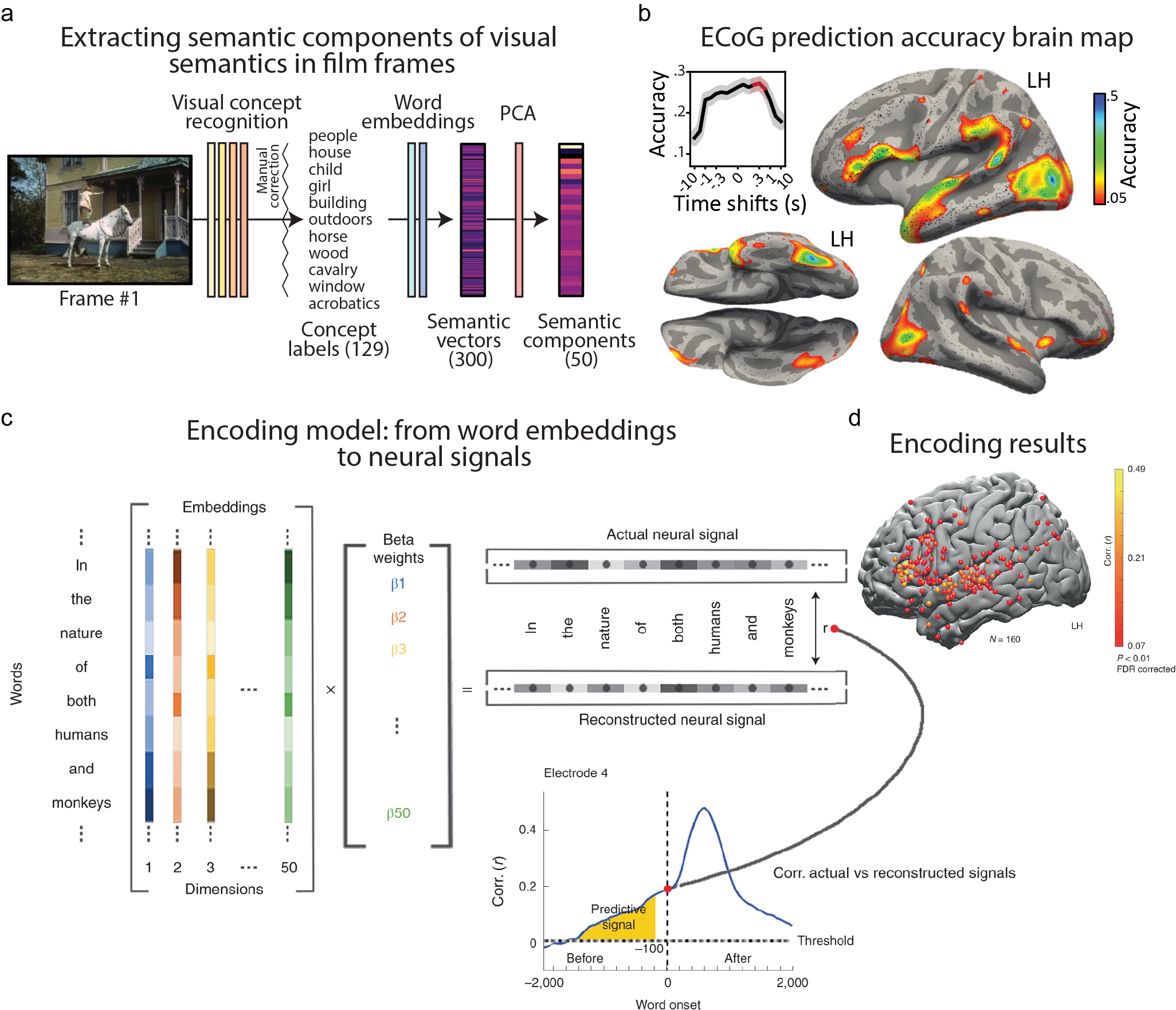}
\caption{\textbf{a-b}. A linear encoding model that predicts ECoG high frequency band activity based on high-level visual features of the movie stimulus~\cite{berezutskaya2020cortical}. Prediction accuracy, measured as correlation between predicted and observed brain activity, is projected from individual electrode center coordinates to a regular grid in common space and interpolated on brain surface. Authors showed that components of visual semantics identified in a data-driven way map onto distinct cortical activation networks. Copyright 2020 Berezutskaya et al. \textbf{c-d}. A linear encoding model that predicts ECoG high frequency band activity during a story listening task~\cite{goldstein2022shared}. Using contextualized word embeddings in their encoding model, the authors demonstrated that neural activity prior to word onset contained information about upcoming words and that this information is used to calculate post-word-onset surprise. Copyright 2022 Goldstein et al.}
\label{figure2}
\end{figure}

Depending on the type of data and problem, various DNN architectures that outline network structure and connections can be used. \emph{Multi-layer perceptrons} (MLPs) are feed-forward fully-connected DNNs as described by Eq.~\ref{eq:1.8}. \emph{Convolutional neural networks} are feed-forward neural networks that use convolutions instead of standard matrix multiplication. They have had tremendous success in computer vision tasks, such as visual object recognition, image colorization and super-resolution. \emph{Recurrent neural networks}~\cite{hochreiter1997long} are cyclic models best suited for learning complex temporal dynamics in the data and have therefore been popular in language processing tasks, such as language modeling, machine translation and speech recognition. Recurrent architectures have recently been outperformed by \emph{transformer} architectures~\cite{vaswani2017attention} -- feed-forward neural networks with an encoder and decoder component and attention modules for efficient input processing. Deep belief networks, autoencoders~\cite{hinton2006reducing}, graph DNNs and other DNN variants also have their practical applications. For a detailed overview of DNN architectures we refer readers to dedicated reviews~\cite{shrestha2019review, khamparia2019systematic, livezey2021deep, abdel2022review}.

Convolutional and recurrent DNNs have been successfully used for automated complex feature extraction. High-level learning tasks, such as object recognition or language modeling, and hierarchical structure of deep neural networks allow extraction of increasingly more complex representations from basic naturalistic inputs, such as images, sounds and texts. The resulting features across neural network layers have been used for linear mapping onto brain activity, first with non-invasive techniques~\cite{yamins2014performance,kell2018task} and more recently using iEEG~\cite{kuzovkin2018activations, berezutskaya2020cortical} (Fig.~\ref{figure2}). Many studies have demonstrated that DNN features explain and correlate with brain activity better than alternative, typically hand-engineered features in language~\cite{schrimpf2021neural, goldstein2022shared}, audio perception~\cite{gucclu2016brains, kell2018task} and vision~\cite{yamins2014performance, cadieu2014deep, cichy2017dynamics} (Fig.~\ref{figure2}).
Some iEEG work explored gradients of feature complexity throughout the model, similar to analogous work in neuroimaging~\cite{kuzovkin2018activations, berezutskaya2020cortical}. 

\subsubsection{End-to-end encoding models using deep learning}

Success of deep learning models in extracting features that accurately predict observed activity throughout the brain, has led to the "deep learning revolution" in cognitive neuroscience. DNN feature models and activations became widely used for exploration and interpretation of the neural mechanisms underlying cognition and perception. Studies identified DNN properties crucial for accurate modeling of observed neural responses, such as hierarchical structure with increasingly complex features~\cite{gucclu2015deep, kell2018task}, local recurrence~\cite{kietzmann2019recurrence, kar2019evidence}, feedback connections, non-linear mixed selectivity of single neurons~\cite{chaisangmongkon2017computing} with ensemble-based feature separability~\cite{parde2021closing}. See~\cite{kriegeskorte2015deep, marblestone2016toward, yamins2016using, grill2018functional, kietzmann2018deep, richards2019deep, sinz2019engineering, kell2019deep, cichy2019deep, hasson2020direct, saxe2021if, lindsay2021convolutional} for perspectives and reviews and~\cite{ullman2016atoms, rajalingham2018large, kay2018principles, zador2019critique, jacob2021qualitative} for critical reports. 

\begin{figure}[!htbp]
\centering
\includegraphics[width=1\linewidth]{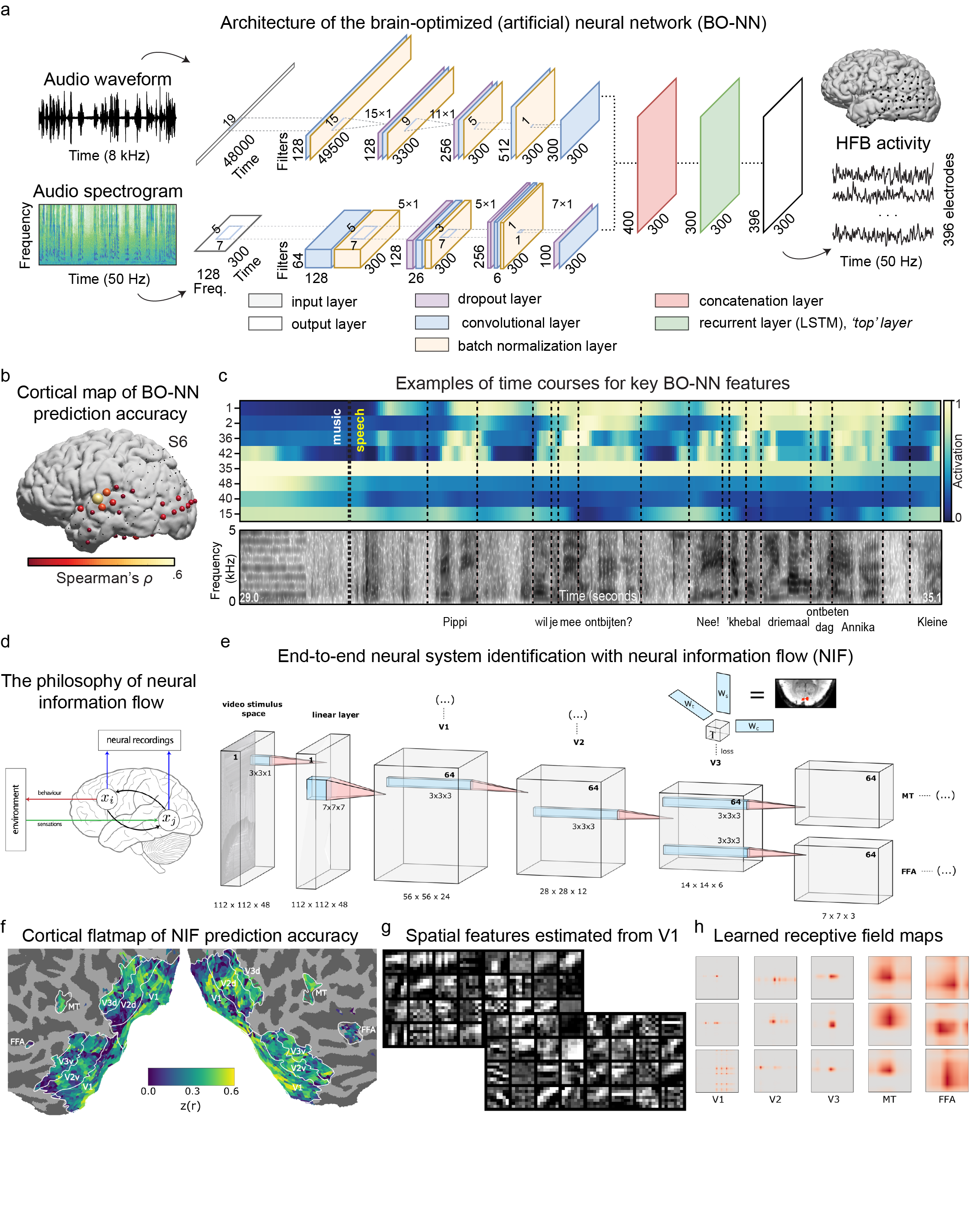}
\caption{\textbf{a-c}. End-to-end brain-optimized encoding model based on a recurrent convolutional neural network (BO-NN)~\cite{berezutskaya2020brain}. The model was trained to predict ECoG responses to a movie soundtrack. Model predictions generalized to a novel movie stimulus watched by a separate group of subjects. Extracted features encoded speech acoustic and temporal information and revealed a gradient of information propagation in the brain during speech perception. Copyright 2020 Berezutskaya et al. \textbf{d-h}. End-to-end encoding model based on neural information flow (NIF)~\cite{seeliger2021end}. The model was trained to fit fMRI responses to a movie stimulus throughout the ventral visual pathway (V1 to MT and FFA). This approach allowed for bottom-up estimation of input visual features that drove responses throughout the cortical hierarchy of visual processing and area-specific receptive fields. Copyright 2021  Seeliger et al.}
\label{figure3}
\end{figure}

Following this line of thinking, several researchers in the field have argued that the next logical step in this direction is development of \emph{end-to-end} deep learning models that map stimuli and behavior to brain activity directly~\cite{kriegeskorte2015deep, saxena2019towards}. Attempts for training and validating end-to-end DNNs that perform non-linear processing of simple input features (such as image pixels or sound waveforms) to directly predict observed brain dynamics have already been made in animal~\cite{joukes2014motion, batty2016multilayer, klindt2017neural, ecker2018rotation, cadena2019deep} and human~\cite{berezutskaya2017modeling, gucclu2017modeling, wen2018neural} research. Recent work applied this approach to iEEG during audio and speech processing~\cite{berezutskaya2020brain, keshishian2020estimating} and extracted data-driven features from input sound data in a way that maximizes model fit to observed neural responses (Fig.~\ref{figure3}). Such end-to-end DNN encoding models explained more variance in neural signals compared to linear encoding models and generalized well to novel experimental data.

The ultimate goal of such models is not limited to novel strategies to extract input features that best explain brain data, but rather providing an accurate computational account of neural processing mechanisms underlying perception and cognition. This remains a challenging task. For example, there has been a discussion of the biological feasibility of current DNN architectures and their learning mechanisms~\cite{marblestone2016toward, lake2017building, kietzmann2018deep, richards2019deep, saxe2021if}. Furthermore, DNNs are often referred to as "black-box models", excellent for learning complex mapping between inputs and outputs, but difficult for interpretation of their intermediate computations~\cite{kay2018principles, grill2018functional, sinz2019engineering, zador2019critique, kell2019deep, cichy2019deep, lindsay2021convolutional, saxe2021if}. It is especially the case for neural modeling as opposed to use of DNNs as feature models, where several visualization and computational tricks have been developed to help interpret what the model has learned~\cite{zeiler2014visualizing, zhou2018interpreting, kietzmann2018deep, barrett2019analyzing, saxe2021if, lindsay2021convolutional}.

Extending upon the previous ideas of end-to-end encoding and incorporating some of the mentioned concerns, the neural information flow (NIF) framework was recently developed~\cite{seeliger2021end}. The framework is based on an end-to-end deep learning model of brain activity, constructed in a modular way, such that each artificial neuronal population (i.e. individual DNN layer), predicts, or projects to, a corresponding biological neuronal population (i.e. region of interest in the brain). In the case of modeling the ventral pathway of visual processing, illustrated by the authors, each convolutional layer (I, II, III, etc) predicted brain responses to a single corresponding brain area along the visual hierarchy (V1, V2, V3, etc). Modifications in model architecture allowed NIFs to explicitly model hemodynamic response function of fMRI (also see previous work on this~\cite{gucclu2017modeling}). The model was validated with fMRI responses to naturalistic visual stimuli (Fig.~\ref{figure3}). It could accurately predict observed brain responses through learning of gradually increasing visual receptive fields, plausible haemodynamic response rates and high-level features of visual semantics. Overall, this end-to-end deep learning framework demonstrates that it could provide data-driven interpretable neural encoding models based on deep learning by incorporating known properties and constraints of neural systems, perceptual processing and brain recording modality. Such models could be used for validation of existing models of neural processing (such as in the ventral visual pathway), model comparison and creation of new, fully data-driven models, whose structure, connectivity and learned perceptual features are optimized to fit observed neural data. 

Following the growing tendency for integration of deep learning in neuroscience and the ever increasing amounts of neural data recorded from parallel neural sites over extended periods of naturalistic stimulation, it is likely that this approach will become dominant in the field. With the help of Bayesian approaches, effective model comparison and hyperparameter optimization can be performed for improving the explanatory power of end-to-end models. 

Guided interpretable designs, including model architecture, objective function and learning algorithm, in the spirit of explainable AI will be key in modeling brain data using end-to-end deep learning models of information encoding. Key principles of neural computation known from previous work may need to be incorporated in the architectures of these models, including hierarchical processing with integrated local and feedback recurrence loops~\cite{douglas2007mapping, bastos2012canonical}, neural adaptation~\cite{noguchi2004temporal, krekelberg2006adaptation, benda2021neural}, sparse coding principles~\cite{olshausen2004sparse}, temporal stability for noise robustness and code invariance~\cite{foldiak1991learning, kording2004complex, clopath2017variance}, stochasticity in neural signals~\cite{fiser2010statistically, orban2016neural} and oscillatory dynamics~\cite{ward2003synchronous, buzsaki2004neuronal}. However, given that detailed workings of many of these principles remain debated in neuroscience, end-to-end DNN encoding models could also provide an excellent framework for testing associated theory-driven hypotheses in silico~\cite{cichy2019deep}. 

Regarding the objective function, end-to-end deep learning models aim to mimic neural processes. Elements of this have been used in representational distance learning - an approach in which DNN feature models were trained with auxiliary objective functions~\cite{mcclure2016representational}. The latter constrained DNN layer activations forcing them to approximate stimulus similarity structure observed in neural responses across brain areas. Similarly, end-to-end models can implement composite objective functions that represent constraints on DNN dynamics and representations.

Gradient descent via backpropagation has long been considered a biologically implausible learning algorithm. However several studies have shown biologically grounded approximations thereof~\cite{guerguiev2017towards, scellier2017equilibrium, whittington2017approximation, sacramento2018dendritic}. Moreover, a lot of promising work is done towards the development of more biologically plausible learning mechanisms similar to backpropagation~\cite{balduzzi2015kickback, lillicrap2016random, pozzi2018biologically, ahmad2020gait}. Moreover, alternatives to error-guided supervised learning such as reward-based learning or unsupervised (Hebbian) learning, are also an active area of research~\cite{movellan1991contrastive, hinton2002training, whittington2017approximation, detorakis2019contrastive}. 

The outlined end-to-end encoding approach using deep learning models is particularly promising in the case of iEEG, as more and more human iEEG data is becoming publicly available~\cite{miller2019library, nejedly2020multicenter, berezutskaya2022open, woolnough2022dataset, peterson2022ajile12, nieto2022thinking}. The long-duration and continuous nature of iEEG recordings makes them an excellent candidate for training and validation of flexible, interpretable end-to-end models of information encoding in the brain, albeit the application of this approach to iEEG is associated with additional iEEG-specific challenges. Such challenges relate to generalization across subjects due to individual coverage and electrode differences, the scale of neural signal representation and use of rate-based approaches as opposed to integration of single neuron activity in spiking neural networks for modeling local field potentials; and overall incomplete understanding of the nature of iEEG signals, its broadband spectrotemporal and oscillatory features. Some of these challenges can be addressed by incorporation of iEEG forward models~\cite{vermaas2020include, vermaas2020femfuns} and attempts at transfer learning in the search for common space across individual subject datasets~\cite{peterson2021generalized}, yet new creative approaches may be required to tackle data complexity. 

Taken together, the application of deep learning models to neuroscience data, and iEEG in particular, remains an exciting and promising direction of research in creating a detailed computational account of neural mechanisms underlying perceptual and cognitive processes.
\\
\\
\subsubsection{Representational similarity analysis (RSA) in iEEG}
Representational Similarity Analysis (RSA) is a multivariate approach, initially proposed by Kriegeskorte and colleagues~\cite{kriegeskorteRepresentationalSimilarityAnalysis2008, kriegeskorteRelatingPopulationCodeRepresentations2009,diedrichsen2017representational}, which focuses on assessing distances between patterns rather than providing decoding accuracy. The principle of RSA is to estimate how the similarity of neural responses to a set of stimuli matches the similarity of perceptual or cognitive evaluations. Distances can also be computed based on predicted responses according to a specific computational model in order to test its fit in distributed regions across the brain or in different behavioral tasks. One of the major advantages of RSA distance matrices is to provide a robust solution to compare results across brain regions, subjects, imagery modalities (e.g., fMRI, EEG, MEG), and even species (i.e., humans, monkeys). As such, RSA method could greatly help integrating experimental research across laboratories and connecting methodological branches of systems neuroscience~\cite{kriegeskorteRepresentationalSimilarityAnalysis2008, schrimpfIntegrativeBenchmarkingAdvance2020,diedrichsen2017representational}.

A handful of studies have combined RSA with the high temporal resolution of iEEG to assess the stability and changes of neural representations during cognitive tasks (see also Chapters 40 and 46). Chang et al.~\cite{changCategoricalSpeechRepresentation2010} first used RSA to highlight the categorical neural representations of speech sounds, mapping their acoustic properties, in the posterior superior temporal gyrus. Zhang et al.~\cite{zhangGammaPowerReductions2015} have characterized time-resolved gamma-band activity patterns in a navigational task, demonstrating the dynamic changes of path representations during encoding and retrieval. More recently, RSA has been used to characterize the dynamics of semantic coding combining current models of semantic memory and neural representations inside the ventral and anterior temporal cortex~\cite{chenWhenWhereSemantic2016, rogersEvidenceDeepDistributed2021}. 

It is also noteworthy that RSA is increasingly used to estimate similarity across biological and artificial neural networks~\cite{yamins2014performance, reddyRepresentationalContentOscillatory2021, rajalingham2018large, bashivanNeuralPopulationControl2019}, enabling testing for nonlinear contributions of features in models~\cite{khaligh-razaviFixedMixedRSA2017}; see~\cite{lindsay2021convolutional} for a review. The combination of AI algorithms and RSA has only been applied to human intracranial EEG data in very few recent vision-related studies. Kuzovkin et al.~\cite{kuzovkin2018activations} demonstrated that visual complexity along the ventral visual pathway, also visible in layers of deep convolutional neural networks (DCNN), was best predicted by gamma activity. Grossman et al.~\cite{grossmanConvergentEvolutionFace2019} revealed face-selective responses in the brain to match the structure of intermediate layers of the DCNN. This research direction is in its early stages but already demonstrates its high potential to evidence key functional principles of the human brain.

\subsection{Decoding models of perception and cognition}

\subsubsection{Supervised machine learning}

Supervised decoding approaches have become common practice in cognitive neuroscience~\cite{glaserRolesSupervisedMachine2019}. They generally involve training a model to classify brain signals into target categories. The latter are assigned “labels” and may reflect distinct groups (e.g. controls and patients), experimental conditions (e.g. familiar and unfamiliar face stimuli) or brain states (e.g. wakefulness,  non-REM sleep and REM sleep). The basic principle is straightforward: The data is first split into “train” and “test” sets (or into train/validate/test sets). The training data and associated labels are  used to train a model for class prediction. The training can in principle be formalized as a data-driven learning process by which a model learns to tune the parameters of a decision function to maximize the correct predictions. Model generalization is then explored by evaluating its predictions on the test data (i.e. samples unseen during training).

One way to ensure that the performance of the classifier is not biased by a lucky (or unlucky) choice of train-test split, the entire train-test procedure is generally repeated multiple times (i.e. cross-validation).
In brain decoding work, the input to the classifier consists of brain variables which are, broadly speaking, of two types:  (1) Hand-crafted features derived from raw brain activity (e.g. spectral power values of iEEG signals). This feature extraction process is generally guided by domain-specific knowledge. This process of selecting, manipulating, and transforming raw data into features that can be used in supervised learning is often referred to as feature engineering.  (2) Original brain signals (such as raw iEEG) can also be used as direct input to the classifier. While shallow learning approaches often involve the use of hand-crafted features, techniques such as deep learning generally take the raw data as input, with little or no preprocessing at all (e.g. continuous iEEG raw time series).

\begin{figure}[!htbp]
\centering
\includegraphics[width=1\linewidth]{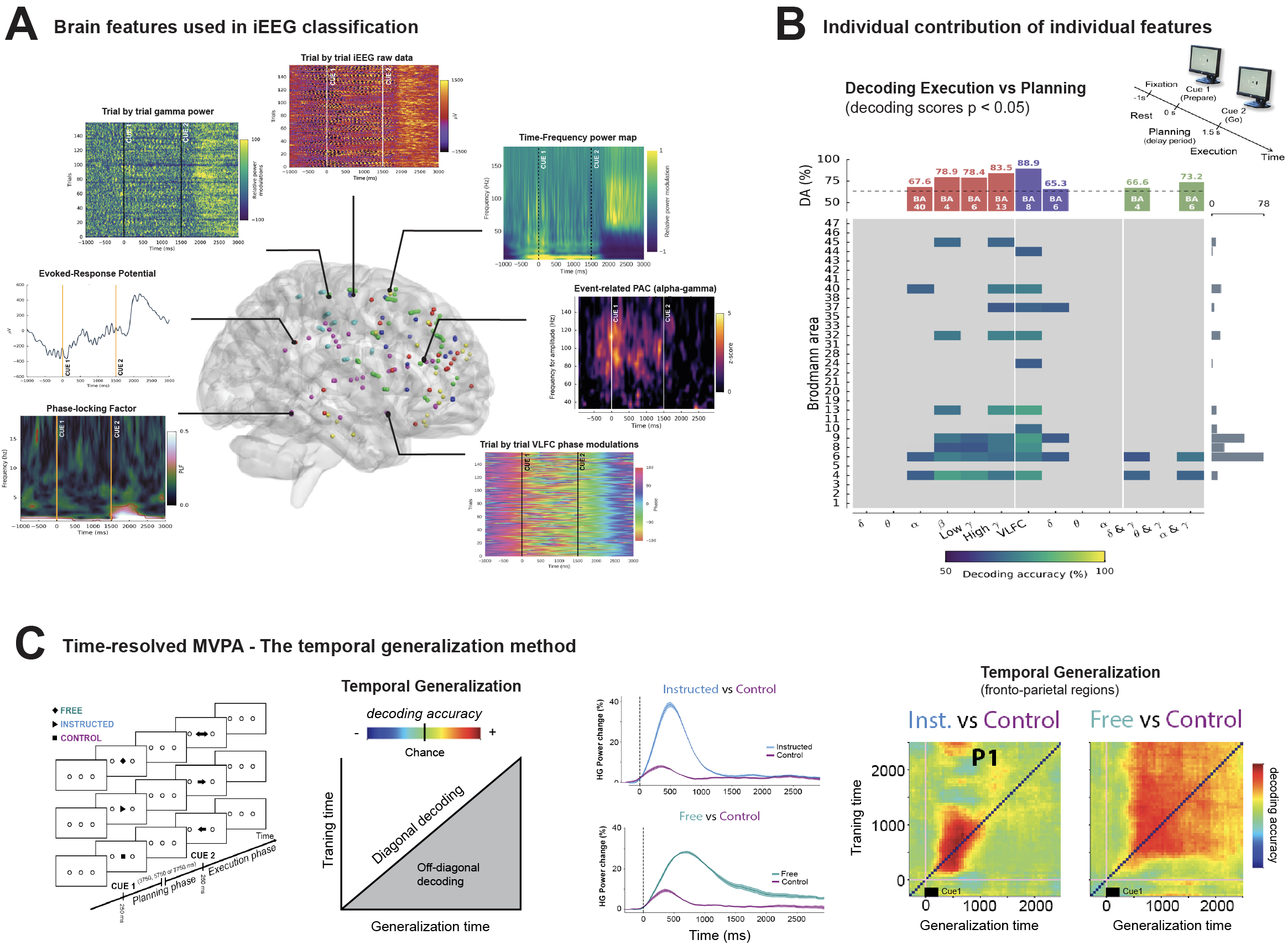}
\caption{\textbf{a-c}. Decoding iEEG data using supervised ML models. A) Examples of brain features computed from iEEG signal and used for classification. B) Decoding accuracies of single-trial classifications of motor-states using power, phase and PAC features in significant iEEG sites. C) Example of temporal generalization of choice-type decoding (Instructed vs Control \& Free vs Control) using high-gamma (HG) activity across significant fronto-parietal iEEG sites. Generalization matrices show decoding performance plotted as a function of training time (vertical axis) and generalization time (horizontal axis). Reprinted with permission from~\cite{combrissonIntentionsActionsNeural2017, thieryDecodingNeuralDynamics2020}.}
\label{figure4}
\end{figure}

A noteworthy issue that often arises when applying supervised machine learning in brain decoding, is the question of how it relates to using classical statistical approaches, such as comparison of means in inferential statistics. An important distinction between the two approaches is that while classical statistics are generally conducted on all the available data, machine learning focuses on out-of-sample generalization (cf~\cite{bzdokClassicalStatisticsStatistical2017} for a more detailed discussion). Furthermore, the use of machine learning cannot be seen as a way to avoid testing the statistical significance of the reported results. In fact, in most cases, an assessment of the reliability of the decoding accuracy of a classifier requires statistical evaluation, for instance using permutation tests~\cite{combrissonExceedingChanceLevel2015} (see also Chapter 44).
\paragraph{Conventional single \& multi-feature classification of iEEG data}
Shallow ML is based on an \textit{a priori} selection of features and aims at identifying the best predictors of distinct cognitive states. In electrophysiology, brain activity has been primarily examined by quantifying increases and decreases in evoked-response potentials and spectral power in distinct frequency bands (Very Low Frequency Component {[0.1–1.5 Hz]}, delta ($\delta$) {[2–4 Hz]}, theta ($\theta$) {[5–7 Hz]}, alpha ($\alpha$) {[8–13 Hz]}, beta ($\beta$) {[13–30 Hz]}, low-gamma (low $\gamma$) {[30–60]} and broadband gamma (high $\gamma$) {[60–200 Hz]}). More recently, other spectral properties such as phase, phase-amplitude coupling, inter-trial coherence, and phase-locking value depicting both local and large-scale neural mechanisms have complemented the features palette used to advance our understanding of the human brain (Fig.~\ref{figure4}a). 

In this context, shallow ML models have proven to be helpful in addressing cognitive neuroscience questions by unraveling the contribution of various intracranial EEG-based features in cognitive processes. More specifically, by computing the decoding accuracy achieved by a specific feature -for example, how well gamma power discriminates between two conditions- we can make interpretations on the importance of this feature in the task. The application of ML as a brain decoding technique is increasingly common in different subfields of systems and cognitive neuroscience including motor control (e.g.,~\cite{combrissonIntentionsActionsNeural2017,wangEnhancingGestureDecoding2020}), sensory perception (e.g.,~\cite{jiangThetaOscillationsRapidly2017a, liuTimingTimingTiming2009,isikWhatChangingWhen2018}), and high-order cognitive processes such as cognitive interference (e.g.,~\cite{vandenieuwenhuijzenDecodingTaskrelevantTaskirrelevant2016}), memory (e.g.,~\cite{familiAutomaticIdentificationSuccessful2017, junPredictionSuccessfulMemory2021}) and decision-making (e.g.,~\cite{thieryDecodingNeuralDynamics2020, terwalHumanStereoEEGRecordings2020}).
\paragraph{ML models}

Various methods have been used to build ML models using intracranial EEG. These include support vector machines (e.g.,~\cite{watrousPhaseamplitudeCouplingSupports2015,hohnePredictionSuccessfulMemory2016, kuzovkin2020identifying}), linear discriminant analyses (e.g.,~\cite{davidescoExemplarSelectivityReflects2014, terwalHumanStereoEEGRecordings2020, thieryDecodingNeuralDynamics2020}), logistic regressions (e.g.,~\cite{whaleyModulationOrthographicDecoding2016, nourskiSpectralOrganizationHuman2014, ezzyatDirectBrainStimulation2017}), and decision-trees such as random-forest (e.g.,~\cite{kuzovkin2020identifying}). The choice of the decoding model impacts the performance and the interpretation of the results (e.g.,~\cite{familiAutomaticIdentificationSuccessful2017}). As all models do not rely on linear algorithms, it is critical to compare their predictions and explore nonlinear or unexpected interactions between variables.
\paragraph{Overview of decoding studies in iEEG}

Using single-feature approach, some researchers have compared how predictive single-trial phase values in rhinal and hippocampal cortices are about encoding memory success, in a word recognition paradigm ~\cite{hohnePredictionSuccessfulMemory2016}). As another example, decoding results have been compared across frontal and parietal cortices during a delayed oculomotor decision task, demonstrating how free-choice in the brain is supported by sustained high-gamma activity during the delay period~\cite{thieryDecodingNeuralDynamics2020}. A similar data-driven approach, has allowed to disentangle the distinct contributions, on a trial-by-trial basis of power, phase, and phase-amplitude coupling to the planning and execution of a goal-directed motor task~\cite{combrissonIntentionsActionsNeural2017} (Fig.~\ref{figure4}b). These studies illustrate how single-feature supervised classification can be employed with iEEG data to advance our understanding of how distinct electrophysiological patterns in individual brain regions are involved in goal directed behavior. In the next section, we will address some promising avenues using multi features ML models.

One important aspect to watch out for when developing ML pipelines is the risk of data overfitting, where the model performs well on training data but does not provide accurate predictions on the previously unseen test data. An overfitted model learns patterns that accurately describe the training data, but do not generalize to the test data. This is a prominent issue in data with small sample size (n) but large number of features, or predictors (p), a situation often referred to as “wide data”, or the “p>>n problem”. This can become an important limitation in intracranial EEG studies where the number of patients is often quite limited whereas the feature space can be very important, which is in principle a perfect recipe for a model to drastically overfit. 

A general recommendation to deal with overfitting is to simplify the model, for instance by decreasing the number of features through feature selection or dimensionality reduction techniques. In intracranial EEG, rather than examining generalization across subjects, intra-subject analyses is often preferred partly because of the heterogeneity of the electrode implantation across individuals. In this case, rather than reflecting the number of participants, the sample size becomes the number of trials, and a new model is trained on each subject using single-trial features with the aim of generalizing across trials, not subjects. This generally reduces the risk of overfitting, assuming there are more trials than features. This also explains why multi-feature classification is not very common in intracranial EEG data sets with modest trial numbers. As a general rule, to reduce the risk of overfitting, it is important to make sure one is not using an overly complex model with too many parameters.

The occurrence of overfitting when applying ML to intracranial EEG data can be detected by cross-validation (i.e. measuring performance on left-out data the algorithm has never seen during training). It is worth noting that sample size directly impacts the standard error of cross-validation, with small sample size studies yielding higher classification performance~\cite{varoquauxCrossvalidationFailureSmall2018}. In addition, with small sample sizes it is particularly important to verify the statistical significance of observed decoding accuracies, as probabilistic chance levels only (e.g. 50\% for binary classification) hold for infinite amounts of data. Large deviations from these thresholds can easily occur by chance in small samples. A recommended approach to handling this issue is to use permutations (i.e. randomly shuffling class labels and recomputing decoding accuracies) to derive a null distribution of classifier performance, and hence thresholds for statistical significance assessments of decoding accuracies~\cite{combrissonExceedingChanceLevel2015} (see Chapter 44 for details). Again, this issue may be particularly important when analyzing small data sets of intracranial EEG. 
\paragraph{Multivariate analyses for iEEG (MVPA, Temporal generalization)}

Among common supervised decoding frameworks, Multivariate Pattern Analysis methods (MVPA) have been attracting growing attention in cognitive neuroscience (for reviews see,~\cite{normanMindreadingMultivoxelPattern2006, haynesPrimerPatternBasedApproaches2015}). MVPA assesses distributed neural representations - also called patterns - across multiple voxels (fMRI) or channels (EEG and MEG) simultaneously. Classifiers are trained to capture the spatial relationships, invisible with traditional univariate tests, between different brain locations across different experimental conditions. Their main advantage is to allow for content-specific decoding in brain activity, and thus to disentangle it from general cognitive functions. In addition, by combining spatial information, MVPA increases signal-to-noise ratio and facilitates single-trial prediction especially relevant in iEEG. Their usefulness has been demonstrated quite extensively in vision, including evidence for sensitivity to orientation in V1~\cite{haynesDecodingMentalStates2006, kamitaniDecodingVisualSubjective2005}, and for distinct pattern of responses for visual categories (e.g., faces, houses) in the ventral temporal cortex~\cite{haxbyDistributedOverlappingRepresentations2001}. A time-resolved application of MVPA called the temporal generalization method has also been introduced (see for a review~\cite{kingCharacterizingDynamicsMental2014}). This method allows for tracking how neural representations unfold in time by training and testing classifiers to discriminate at least two experimental conditions at all points in time~\cite{kingTwoDistinctDynamic2014, stokesDynamicCodingCognitive2013}. This approach allows for detecting and comparing periods of optimal decodability across time and cognitive operations.
\\
\paragraph{Overview of iEEG decoding studies using MVPA and temporal generalization}

The application of multivariate analysis techniques to iEEG is still in its early days, though the high spatial and temporal resolutions of these signals would constitute a true advantage to discriminate neural patterns. Tsuchiya and colleagues~\cite{tsuchiyaDecodingFaceInformation2008} have applied multivariate models to ECoG power signals, finding reliable distinct representations for happy and fearful faces in the ventral temporal cortex. Spatial pattern analyses have also been adopted to identify successful word encoding in ECoG signals~\cite{junPredictionSuccessfulMemory2021, gonzalezElectrocorticographyRevealsTemporal2015} and iEEG~\cite{ezzyatDirectBrainStimulation2017}, and to show the reinstatement of similar processes during successful encoding and retrieval of words in iEEG~\cite{kragelSimilarPatternsNeural2017}. Using the temporal decoding method, Thiery et al.~\cite{thieryDecodingNeuralDynamics2020} have recently provided the first evidence for delayed high-gamma activity in the frontoparietal cortex in mediating free-choices compared to instructed choices (Fig.~\ref{figure4}c).

Despite the significant advances that multivariate ML methods allow, certain limitations should be considered when interpreting classification results. Neural patterns obtained from combining depth-electrodes can strongly vary between patients and often cover distant brain regions that may not be anatomically connected. This raises the question about the actual computational usage and biological relevance of the information included in the analysis. In addition, MVPA results do not provide weights or importance of individual electrodes included in the analysis. It is then possible that only a small part of the electrodes contributes to the significant decoding results, or that significant decoding gets lost when too many noisy electrodes are added to the analysis.

\subsubsection{Deep learning for decoding} 

Deep learning is increasingly used in neuroscience research~\cite{Richards2019ADL}. Although still in its early days, the application of deep learning to iEEG is very promising and is starting to gain momentum (see also Chapter 46). Using deep learning in the context of iEEG decoding work is in part appealing because it allows us to go beyond investigation based on handcrafted features (see Section~\ref{subsec:feature_eng}). Indeed, representation learning~\cite{BengioRepLearning2014} is a major motivation for using deep neural networks. In principle, representation learning (or feature learning) is the process by which a system automatically uncovers the representations needed for classification directly from the raw data. Although the features learned through deep learning are often described as being abstract, interpretation can be facilitated using an array of feature visualization techniques~\cite{GrunTaxonVisFeat2016} including deconvolutional methods such as guided back-propagation~\cite{Springenberg2015StrivingFS}. The identification of the most discriminative samples through these approaches can be combined with signal processing tools commonly used in iEEG work, such as spectral analyses, to further enhance interpretability.
\\
\\
An interesting illustration of how deep learning can be applied to iEEG was provided by a study that used deep convolutional neural networks (CNNs) to probe which frequency bands of the iEEG data are correlated with feature transformations of ascending complexity along the ventral visual pathway during object recognition~\cite{kuzovkin2018activations}. The study revealed that gamma activity (30-70 Hz) reflects the increasing complexity of visual feature representations in the deep CNN (see also~\cite{berezutskaya2020cortical}). These results illustrate how CNN activity may capture essential electrophysiological features of biological object recognition not only in space and time, but also in the frequency domain. 

Recent work on decoding in cognitive neuroscience has shown the potential of other deep learning architectures, such as recurrent neural networks~\cite{li2019interpretable}, generative adversarial neural networks~\cite{seeliger2018generative, le2021brain2pix, dado2022hyperrealistic} and transformers~\cite{malkiel2021pre} to reconstruct stimulus information based on non-linear transformations of brain inputs. In decoding from iEEG data, deep learning models have been used as powerful feature extractors from raw iEEG signals and as a tool for combining data across subjects with varying implantation coverage~\cite{peterson2021generalized}. Deep learning has also been applied to iEEG in the context of epilepsy, including seizures detection, e.g. ~\cite{aboujaoudeDetectionMesialTemporal2020, aboujaoudeNoninvasiveDetectionHippocampal2022, ahmedt-aristizabalAutomatedAnalysisSeizure2017,antoniadesDetectionInterictalDischarges2017,antoniadesDeepNeuralArchitectures2018, burrelloHyperdimensionalComputingLocal2020, caffariniEngineeringNonlinearEpileptic2022,chungDeepConvolutionalNeural2020,yamamotoDatadrivenElectrophysiologicalFeature2021} (see Section~\ref{subsec:dbs}) as well as for brain-computer interfaces~\cite{hashimotoSwallowingDecoderBased2021,pradeepkumarDecodingHandGestures2021,sliwowski2022decoding,xieDecodingFingerTrajectory2018} (see Section~\ref{sec:technology}).

\section{AI-iEEG for neurotechnology} 
\label{sec:technology}


The ever growing amount of neuroscientific knowledge has gradually made it possible for a novel field of neurotechnology to emerge. This field is focused on development of brain-computer interface (BCI) devices, and one of its fundamental goals is to offer technological solutions for real-world clinical problems. This includes technology for 1) restoration of lost cognitive and motor functions, such as visual prosthesis for the blind, BCIs for severely paralyzed individuals or cochlear implants for people with impaired hearing (see also Chapter 60); and 2) therapy and treatment of various chronic neurological conditions, such as deep brain stimulation devices for the Parkinson's disease, dystonia, epilepsy and severe mental disorders (see also Chapters 59 and 61). Despite the availability and lower safety concerns of non-invasive brain recording techniques, intracranial technology has proven to be best suited for effective management of severe neurological conditions and restoration of severed cognitive or motor functions. This is because it provides superior neural signal quality~\cite{ball2009signal}, allows access to highly localized surface and deep brain structures, and overall has a better potential for long-term, continuous brain recordings and autonomous home use of the device. In that regard, ECoG and sEEG recordings have been instrumental in development, testing and deployment of various examples of emerging clinical neurotechnology. See~\cite{wolpaw2002brain, birbaumer2007brain, leuthardt2009evolution, moran2010evolution, shih2012brain, chakrabarti2015progress, martin2019use, herff2020potential, miller2020current, muller2021invasive, saha2021progress, rapeaux2021implantable} for reviews. 

Neurotechnology is a highly multidisciplinary field with contributions from neuroscience, signal and data processing and electrical engineering disciplines. Fundamental neuroscience provides the theory of neural processes that support cognition and behavior, and the understanding of neural signals in health and disease. Electrical engineering and material science develop cutting-edge neural recording and stimulation technology and necessary hardware to enable operation and power supply of the device. Statistics, computational modeling and machine learning provide the methodology and algorithmic base for processing neural signals and addressing the clinical problem at hand, via control of external devices (a wheelchair, exoskeleton or computer) or brain stimulation to alleviate symptoms of a chronic neural condition. More recently, BCI methodology began to incorporate advanced computational models that make use of machine learning and deep learning approaches. These AI-powered algorithms are being used increasingly more often for 1) neural signal preprocessing and extraction of informative temporal and spectral features in brain data; 2) decoding models of target neural events, such as intended movements in paralyzed individuals or biomarkers of pathological activity in individuals with a neurological condition; 3) optimization of BCI algorithms and development of energy-efficient computing systems; 4) integration with external domain-specific tools and applications, such as virtual reality, natural language processing models and robotic components, for boosting the development of cutting-edge BCI solutions. 

In this section we will describe neurotechnology applications based on the combination of AI methodology and iEEG neural data. The main examples will include BCIs for speech and communication in severely paralyzed individuals, brain implants for motor control of a wheelchair, exoskeleton or robotic arm, and deep brain stimulation devices for neurological and psychiatric disorders. 

\subsection{IEEG BCI for speech and communication}

Individuals who suffer from a motor neuron disease (such as amyotrophic lateral sclerosis) or a brainstem stroke can develop severe motor paralysis. In extreme cases it can lead to a "locked-in syndrome" (LIS)~\cite{workgroup1995recommendations} and result in a complete loss of muscle control. Such individuals may lose the ability to perform voluntary body movements, such as walking, object interaction, speech, facial expressions, blinking, swallowing and, in extreme cases, breathing. Their means of communication are often reduced to limited eye movements or residual control of a few facial muscles. People with a complete LIS may lose any ability to communicate. Given modern standards of healthcare "locked-in" individuals can survive in their severely paralyzed state for many years and even decades~\cite{Doble2003,laureys2005locked}. Interestingly, studies report that they can experience good quality of life~\cite{robbins2001quality, neudert2004individual, lule2008depression, lule2009life, linse2017eye}, and that one of the major predictors of good quality of life is retaining the ability to communicate with the outside world~\cite{bruno2011survey, rousseau2015quality}.

In many cases, motor paralysis in people with LIS is caused by damage to connections between the brain motor cortex and the spinal cord, or the spinal cord and nerves that lead to the muscle tissue. Unlike patients in a vegetative state, "locked-in" individuals remain conscious. Cortical activity in individuals who have suffered a brainstem stroke is largely unaffected and exhibits patterns similar to that of the healthy able-bodied individuals. Effects of motor neuron degeneration on cortical activity in individuals with a motor neuron disease, such as ALS, remain less well-understood. Several studies report a decrease in alpha~\cite{santhosh2005decreased} and theta~\cite{babiloni2010resting} power compared to healthy controls, while other studies report an increase in alpha power~\cite{iyer2015functional} or no difference~\cite{geronimo2016performance}. A recent study indicates that population-level sensorimotor dynamics in ALS subjects may be comparable to sensorimotor responses in non-human primates~\cite{pandarinath2015neural}. Consistent with this finding, several studies reported successful decoding of motor information from sensorimotor cortex of individuals with ALS~\cite{vansteensel2016fully, oxley2021motor}. This work demonstrates the potential of BCI technology to detect intended communication signals from brain activity of "locked-in" individuals and translate them to computer commands, thereby unlocking a means of communication with the world.

Preparatory research towards BCI technology based on an iEEG implant relies on pre-clinical studies in able-bodied patients with medication-resistant epilepsy. These patients are temporarily implanted with iEEG electrodes (typically, for 7-10 days) for clinical monitoring of their condition with a goal to identify and subsequently remove neural sources of epilepsy. While implanted with iEEG, such patients can participate in various cognitive and motor tasks, and the collected iEEG signals are subsequently analyzed with a goal to identify, or decode, neural events relevant for BCI research. Most of this work rests on the assumption that speech production and other motor activity in able-bodied subjects engage brain mechanisms and cortical areas similar to those of attempted speech and imagined actions of paralyzed individuals. This assumption is supported by the recent work on attempted movements in amputees~\cite{kikkert2016revealing, bruurmijn2017preservation} and paralyzed individuals~\cite{guenther2009wireless, freudenburg2019sensorimotor}. Given the reports on potentially shared neural basis of performed and imagined movements, including speech~\cite{leuthardt2011using, pei2011decoding, martin2014decoding, brumberg2016spatio, angrick2021real}, it has been proposed to use imagined movement paradigms in able-bodies participants as a proxy for development of BCI technology for individuals with LIS. Other reports, however, provide conflicting evidence regarding the shared neural mechanisms of imagined and performed motor activity~\cite{hermes2011functional, maegherman2019motor}, and in general, a distinction between attempted (no actual movement possible) and imagined (actual movement possible and likely inhibited) movement should be made. Further research comparing neural activity during performed, imagined and attempted actions is needed to gain better understanding of these processes and inform BCI development. 

The pre-clinical work on iEEG BCIs has been focused on identifying three key components of the emerging technology for communication: 1) optimal location for implanting iEEG electrodes, 2) optimal targets or features for decoding, and 3) optimal neural decoding model. State-of-the-art BCI technology for long-term autonomous use is limited to implants with a small number of electrodes (four or sixteen)~\cite{vansteensel2016fully, oxley2021motor}. Larger numbers will require higher power consumption for continuous signal recording and analysis and may lead to overheating of the implanted components beyond reasonable temperatures. Moreover, implantation of electrodes over large parts of the cortex may lead to higher risks during and following the implantation surgery and may result in longer recovery times. Until these concerns are mitigated, targeting a smaller brain region for BCI implantation is preferred. Currently, sensorimotor cortex is the primary candidate for such BCIs as its involvement in muscle control during movement and communication has been studied extensively with iEEG in non-human primates~\cite{shin2012prediction, chen2013prediction, umeda2019decoding} and human participants~\cite{pulvermuller2006motor, bouchard2013functional, branco2017decoding, chartier2018encoding}. Other brain areas have also been considered in decoding of language and speech, such as inferior frontal gyrus, temporal and parietal regions~\cite{kellis2010decoding, pei2011decoding, pei2011spatiotemporal, ikeda2014neural, kanas2014joint, martin2014decoding, herff2015brain, martin2016word, brumberg2016spatio}. These regions are part of the language processing network in the brain~\cite{hickok2007cortical, friederici2011brain}. However, language-related neural signals tend to be highly spatially distributed and varied across subjects, which results in less overall consensus about their potential for BCI use. It remains to be seen how the development of BCIs for communication can incorporate current neurolinguistic theory and benefit from signals recorded in the language processing network of the brain. 

Another key component of BCI for communication is the target of decoding, and given the focus on sensorimotor cortex, one of the prominent lines of research is decoding of motor information. This work includes decoding of performed movements, imagined or attempted movements of, for example, the upper limb, and relies on the previously mentioned assumption about the shared neural basis of performed and attempted movements. This research includes decoding of various types of movement, such as simple hand and finger movements~\cite{bundy2016decoding, sliwowski2022decoding}, gestures~\cite{bleichner2016give, branco2017decoding, li2017gesture, pan2018rapid, verwoert2021decoding} and elements of sign language~\cite{leonard2020cortical}. Accurate decoding of several classes of facial movements, such as basic mouth movements~\cite{salari2019classification} and facial expressions~\cite{salari2020classification} has also been demonstrated. 

A long-term goal of approaches based on accurate decoding of several discrete motor commands, is to enable communication via development of a BCI with several degrees of freedom for flexible cursor control~\cite{blakely2009robust, leuthardt2011using} or use of a computer-based language speller~\cite{cecotti2011spelling}. Decoded motor commands could then be used to control such interface to move the cursor (in a discrete: up, down, left, right, or continuous fashion) and perform item selection. Despite the potential of this approach, most recent work has been focused on making BCI communication faster and more efficient, and rely on a more natural way of communication, such as attempted speech. For this, BCI researchers have turned to decoding of discrete speech-related features, such as individual phonemes~\cite{blakely2008localization, pei2011decoding, mugler2014direct, bouchard2014neural, ramsey2018decoding}, syllables~\cite{livezey2019deep}, words~\cite{kellis2010decoding, martin2016word} and sentences~\cite{zhang2012spoken}. Researchers have also used a closed set of decoded elements, such as phonemes, as building bricks for potential open-vocabulary decoding of full words and sentences~\cite{herff2015brain, herff2019generating}. Some work explored decoding of acoustic features of speech~\cite{pasley2012reconstructing,martin2014decoding} and used external AI models to synthesize speech from them~\cite{angrick2021real, kohler2021synthesizing}. Another recent study used microelectrode arrays over the hand knob in sensorimotor cortex to decode handwriting patterns during word spelling~\cite{willett2021high}. Decoding of speech and speech-related features has proven to be more difficult compared to other motor movements. On the other hand, it can offer the potential to provide a faster, more natural and convenient way to communicate. Therefore, speech and speech-related features remain a more attractive and overall preferred target of decoding in BCI research.

The third component -- optimal models for decoding -- also remains a topic of debate. From the beginning, BCI development has relied on various statistical and machine learning algorithms to perform signal processing, dimensionality reduction, time series analysis and decoding~\cite{van2009brain, nicolas2012brain}. Various classification methods, such as logistic regression, support vector machines, k-nearest neighbours, random forests, artificial neural networks and others have been used and compared in terms of their decoding accuracy~\cite{brumberg2011classification, kanas2014joint, ramsey2018decoding, livezey2019deep, anumanchipalli2019speech}. There have been attempts to boost performance by aggregating results from an ensemble of classifiers, each generating its own decoding output. Latest advances in AI have led to the focus on the so-called  \emph{neuroengineering} approach that relies on deep learning for decoding. This includes the use of deep learning models for extracting features that could be used as targets of decoding~\cite{akbari2019towards}, as well as end-to-end deep learning models for learning a complex mapping from neural data to speech and language features~\cite{anumanchipalli2019speech,akbari2019towards,livezey2019deep,makin2020machine,sun2020brain2char,angrick2021real,kohler2021synthesizing}. Deep learning has been used for transfer learning in iEEG to account for the variability in electrode placement across subjects and pre-train decoding models for new subjects on previously collected datasets~\cite{makin2020machine, peterson2021generalized}. Another recent trend is the use of external AI tools for boosting the decoding results, such as language models~\cite{moses2016neural, moses2021neuroprosthesis}, audio synthesis models~\cite{akbari2019towards, angrick2021real, kohler2021synthesizing}, speech-to-text models~\cite{makin2020machine} and models of articulatory-to-acoustic inversion~\cite{anumanchipalli2019speech}. Overall, it appears that AI-powered tools for decoding are beginning to dominate the BCI scene due to their superior accuracy and potential for sophisticated data-driven solutions.

Several human clinical trials of iEEG-based BCIs for communication aiming to validate the envisioned technology in the daily life of end users have already started. A recent study by Vansteensel et al. validated a fully-implantable BCI device for communication in an individual with late-stage ALS~\cite{vansteensel2016fully}. Using a four-channel ECoG implant over the subject's sensorimotor cortex and a simple linear decoding model the authors decoded binary "click" events based on attempted hand movements with 90\% accuracy. The decoded clicks were used to control the on-screen computer menu and language speller. For the first time, an implantable BCI technology for communication offered its user the possibility of autonomous 24/7 home use. More recently, another study reported preliminary data from a fully-implantable BCI system in two ALS individuals~\cite{oxley2021motor}. Minimally invasive stent-electrodes (Stentrode, Synchron, CA, USA)~\cite{oxley2016minimally} were implanted within participants' vein adjacent to their motor cortex. The study reported high-fidelity decoding of two and three classes of motor events (on average, 93\% accuracy of binary decoding in the typing task) and thereby demonstrated high potential of their BCI system for autonomous use in daily life. Another recent study reported the proof-of-concept of decoding attempted speech using advanced AI models in a patient with anarthria (inability to speak) following a brainstem stroke~\cite{moses2021neuroprosthesis}. The authors implanted a 128-channel ECoG grid (not suitable for 24/7 home use due to power limits) over sensorimotor cortex, and built a deep recurrent neural network to detect attempted speech in neural signals in real time. After detection, another set of deep neural networks decoded individual words out of a closed set of 50 words. In addition, the system used a hidden Markov model approach and an external language model to decode short full sentences made of the set of 50 words. The study reported highly accurate (word error rate of 26\%) decoding results, highlighted the importance of using external language models in achieving this result, and concluded that the developed BCI was an unprecedentedly fast (15 words a minute), accurate and naturalistic means of communication available for their participant.

\begin{figure}[h]
\centering
\includegraphics[width=1\linewidth]{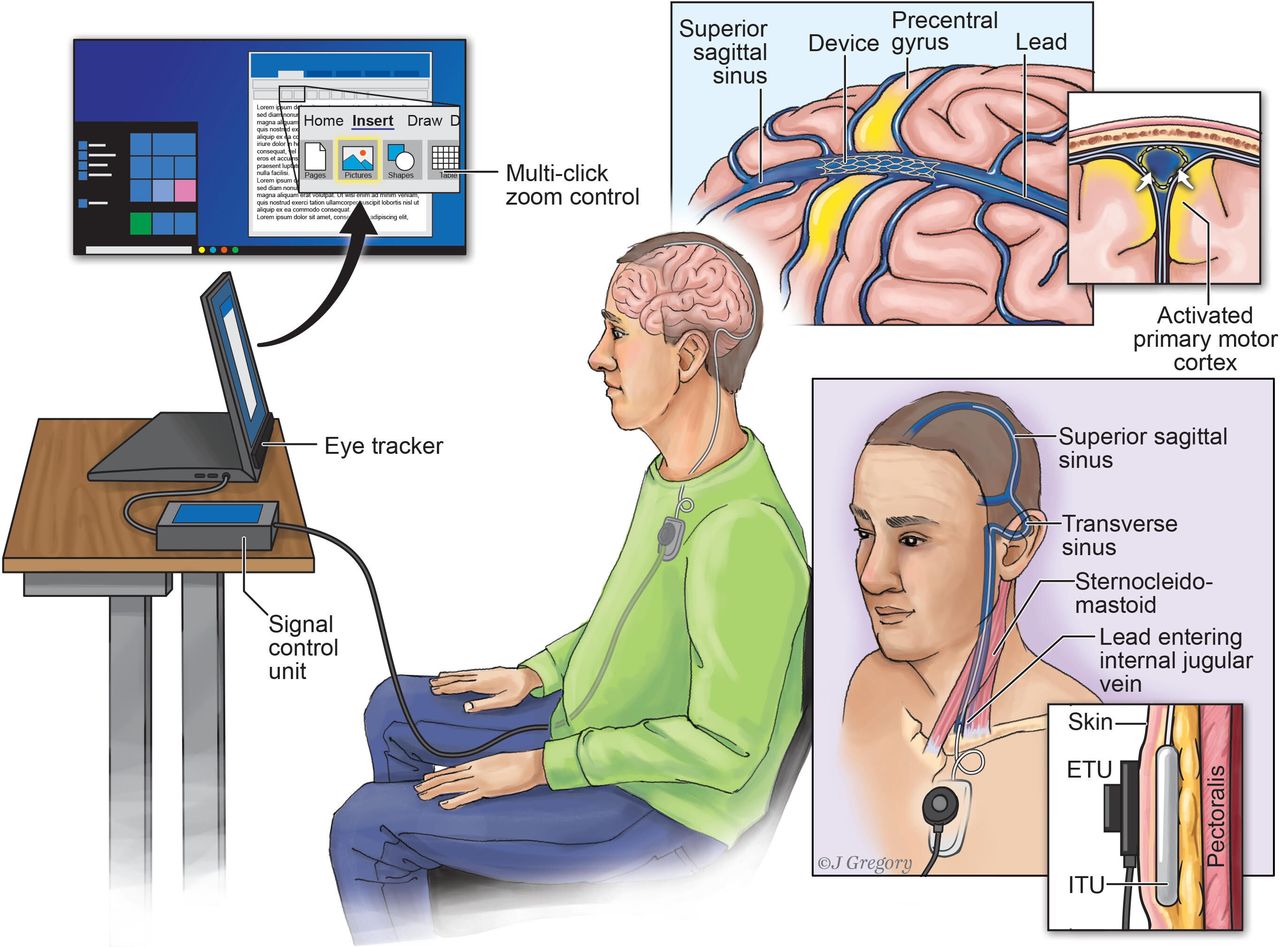}
\caption{Stentrode-based BCI for multi-click computer control used by an individual with ALS. The intracranial electrode array is implanted within the participant's jugular vein. The study reports a successful case of autonomous home use for a BCI technology that provides a means of communication for individuals with severe motor paralysis~\cite{oxley2021motor}. Reprinted with permission from BMJ Publishing Group.}
\label{figure5}
\end{figure}

Currently this work remains at the stage of clinical trials and research, but given its promising results and continuous progress in iEEG technology, BCI hardware, signal processing and decoding models, it may achieve commercialization in the foreseeable future. This will make BCI-based communication technology available for "locked-in" individuals. Once this technology becomes part of the healthcare system, it will allow its users to communicate more efficiently, thereby helping them re-integrate in the society and further improving the quality of their lives.

\subsection{IEEG BCI for motor control}

Up to 50\% of individuals who undergo a spinal cord injury can develop a condition called tetraplegia, which refers to a complete paralysis of four limbs~\cite{anderson2004targeting}. BCI for restoration of motor control aims to improve the lives of these individuals by decoding intended movements from neural signals in their sensorimotor cortex, and use them to control an external device. The latter can be a cursor on a screen, a virtual reality avatar or a mobility device, such as a robotic arm, exoskeleton or wheelchair. As this neurotechnology is focused on restoration of a lost function, its target population, goals and approach overlap with those of the BCIs for communication. Yet despite the severe paralysis, individuals with tetraplegia retain their ability to communicate, and this BCI technology aims to further decrease the reliance of its users on patient care by allowing them to move and interact with objects independently. Since these individuals suffer from paralysis of all four limbs, and because they are not likely to recover from this condition, long-term invasive BCI solutions may be best for restoring  these persons' motor control.

The field of iEEG-based BCI for motor control has been based on work on single cell recordings and motor-related action potentials, and has therefore been dominated by microelectrode Utah arrays. Successful preliminary work on movement decoding in animals~\cite{serruya2002instant, taylor2002direct, carmena2003learning, paninski2004spatiotemporal, velliste2008cortical, moritz2008direct} has laid down the pathway for clinical trials in humans (see Chapter 60). In 2006, first decoding results using a microelectrode Utah array implanted in a tetraplegic individual were published~\cite{hochberg2006neuronal}. The study participant learned to control a cursor using attempted hand movements. In each trial the participant was cued with one of four target positions (up, down, left and right) and was asked to attempt to move the cursor from the center of the screen to that position. The associated single cell activity on the motor cortex showed modulation that was picked up by the decoding algorithm and used to update the cursor position. The participant reached accuracy of 73–95\% across six sessions. This study was part of the first clinical trial BrainGate (Cyberkinetics, Inc.) aimed at development of microelectrode-based BCIs for motor control in paralysed human subjects (\hyperlink{https://clinicaltrials.gov/ct2/show/NCT00912041}{BrainGate}). 

Later studies, including work from other groups (\hyperlink{https://clinicaltrials.gov/ct2/show/NCT01364480}{University of Pittsburg} and \hyperlink{https://clinicaltrials.gov/ct2/show/NCT01997125}{Ohio/Neurolife}), expanded upon these results and demonstrated microelectrode BCI in tetraplegic subjects for a real-time control of various external devices: computer cursor~\cite{simeral2011neural, gilja2015clinical}, robotic arm~\cite{hochberg2012reach, collinger2013high}, driving and flight simulators~\cite{kryger2017flight, dunlap2019towards}. These studies rely on decoding of a few discrete motor commands, for example, up, down, left, right~\cite{hochberg2006neuronal}; complex multi-joint movements, such as reach and grasp~\cite{hochberg2012reach, downey2018implicit}; various movement parameters, such as velocity, translation, orientation, torque, etc.~\cite{collinger2013high} and continuous muscle activity~\cite{bouton2016restoring}. Several studies explored the possibility of restoring proprioception via electrical stimulation of the implant~\cite{flesher2016intracortical, ganzer2020restoring} and showed that implementing sensory feedback from interaction of the BCI-controlled robotic arm with objects results in faster, more accurate and naturalistic motor control~\cite{flesher2021brain}. Several studies combined microelectrode BCI with functional electrical stimulation~\cite{ethier2015brain} of peripheral nerves to reanimate participant's paralysed limb~\cite{bouton2016restoring, ajiboye2017restoration, schwemmer2018meeting}. See~\cite{bockbrader2019upper} for an overview of the state of the art in upper limb decoding for BCI motor control.

Despite promising results achieved with microelectrodes, this invasive technology suffers from signal decay over time~\cite{colachis2021long}. It has also been reported that for chronic long-term use of the implant, repeated calibration sessions are required~\cite{silversmith2021plug,pandarinath2017high}. Moreover, microelectrode devices have not been thoroughly tested for autonomous home use (although see some preliminary recent work~\cite{simeral2021home}) and are not currently certified for indefinite implantation, which largely affects and restricts the user experience. As a less invasive technique, iEEG does not suffer from signal decay over long periods of time as strongly as microelectrode arrays do. Interestingly, some work in non-human primates indicates that decoding from iEEG may be more accurate compared to microelectrode arrays~\cite{kanth2020electrocorticogram}. Moreover, iEEG shows less signal instability over time and therefore does not require frequent re-calibration~\cite{blakely2009robust, chao2010long, vansteensel2016fully, degenhart2016histological, larzabal2021long, silversmith2021plug}. Autonomous home use of iEEG implants in human subjects has been successfully demonstrated~\cite{vansteensel2016fully}, and iEEG electrode technology and hardware for power supply continue to evolve towards providing even better long-term recording solutions~\cite{mestais2014wimagine, alahi2021recent}. Altogether, these factors contribute to the emergence of iEEG-based BCI technology for restoration of motor control.

Unlike speech, motor control via implanted neurotechnology can be assessed in animals, and several previous studies have showed promising results of decoding performed and attempted movements in non-human primates implanted with ECoG grids~\cite{chao2010long, watanabe2012reconstruction, shimoda2012decoding, hu2018decoding}.
At the same time, similar to BCI for communication, much of initial research on decoding motor information from neural activity has been done pre-clinically, in able-bodied patients temporarily implanted with iEEG for epilepsy monitoring. This work reported on successful decoding of discrete movements and postures of hand~\cite{toro1994event, levine2000direct, mehring2004decoding, leuthardt2004brain, pistohl2012decoding}, fingers~\cite{kubanek2009decoding, miller2009decoupling, hotson2016individual}, foot~\cite{toro1994event, satow2003distinct} and other body parts; as well as decoding of continuous movement trajectories~\cite{schalk2007decoding, pistohl2008prediction, gunduz2009mapping, kellis2012decoding, nakanishi2013prediction} and movement properties, such as force, velocity, direction and speech~\cite{anderson2012electrocorticographic}, decoding of complex multi-joint movements and gestures~\cite{bleichner2016give, branco2017decoding, li2017gesture, pan2018rapid, verwoert2021decoding}. See~\cite{volkova2019decoding, branco2019encoding, chandrasekaran2021historical} for reviews. Most of iEEG studies on motor decoding to date have focused on ECoG. However, some recent work in able-bodied subjects demonstrates the potential of sEEG in development of BCIs for motor restoration. These studies report successful cursor control~\cite{vadera2013stereoelectroencephalography, murphy2016contributions} and control of a robotic arm~\cite{li2017preliminary} using motor signals decoded from sEEG. See~\cite{herff2020potential, chandrasekaran2021historical} for reviews of pre-clinical work with sEEG. 

Clinical trials in individuals with tetraplegia tested the possibility of real-time clinical applications of ECoG-based BCIs for motor control. Work by Wang and colleagues was among the earlier studies to validate this ECoG-based decoding of intended motor commands in a human subject with tetraplegia~\cite{wang2013electrocorticographic} (but see also~\cite{marquez2009control, yanagisawa2011real}). Following a spinal cord injury, the study participant lost volitional control of arm and hand. The participant was subsequently implanted with a high-density ECoG grid over the sensorimotor cortex and trained to perform 2D control (x and y coordinates) of a computer cursor using attempted hand and elbow movements (thumb for left, elbow for right, both thumb and elbow for up and neither for down). After 11 days of training, in their final session the participant reached an accuracy of 87\% (chance was 8\%) over 176 cursor control trials. Broadband neural activity in high frequency band (gamma and high gamma) showed largest modulation by the task.
In another study, Silversmith et al.~\cite{silversmith2021plug} tested a cursor control BCI system in a patient with severe tetraparesis, or partial tetraplegia. A closed-loop BCI system with a high-density ECoG implant was built to process motor neural signals (imagined arm and head movements) and use them for control of a cursor on screen. The system achieved high performance and stability over time without the need of re-calibration. 

\begin{figure}[ht]
\centering
\includegraphics[width=1\linewidth]{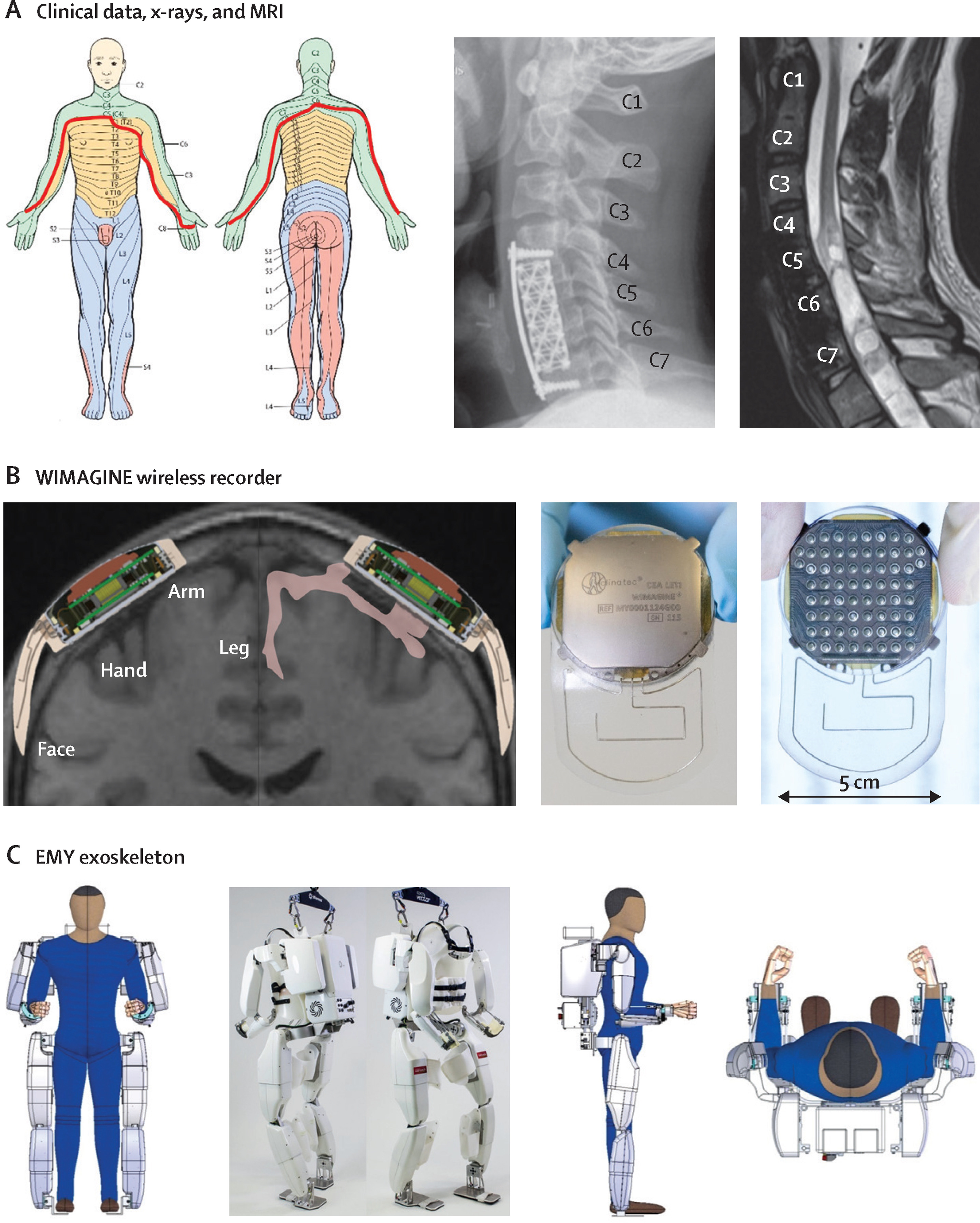}
\caption{ECoG-controlled exoskeleton: a proof of concept~\cite{benabid2019exoskeleton}, pp 1112–1122, reprinted with permission from Elsevier.}
\label{figure6}
\end{figure}

A series of recent studies demonstrated the use of an ECoG-based BCI for control of an exoskeleton~\cite{benabid2019exoskeleton,larzabal2021long, moly2022adaptive}. Benabid and colleagues for the first time showed a wireless ECoG device WIMAGINE~\cite{mestais2014wimagine} implanted epidurally and bilaterally over the sensorimotor cortex of a patient with tetraplegia (Fig.~\ref{figure6}). In two years of training, the implanted individual learned to walk and perform complex upper limb movements by controlling a four-limb exoskeleton or a virtual avatar (virtual reality) with the ECoG implant. Motor control was achieved with an adaptive linear model that decoded movement parameters to the target location (distance to its 3D position and angles between current and target location). After training the decoder, the participant could perform voluntary movement control with a success rate above 70\% for walking and 71\% for 3D bimanual reach and touch tasks. This performance was stable without the need for frequent re-calibration after 7 weeks~\cite{benabid2019exoskeleton} and later than 6 months~\cite{moly2022adaptive} after implantation. 

Another recent study used an ECoG implant in combination with a functional electrical stimulation device to enable cortical control of the paralyzed hand~\cite{cajigas2021implantable}. Through training the study participant learned to trigger electrical stimulation of his hand with 89\% accuracy and successfully perform hand grasping tasks. After the initial period of in-lab testing, the participant was able to use the developed BCI system autonomously at home.

Similar to BCI for communication, restoration of motor control relies on various decoding models from statistics and machine learning (see ~\cite{volkova2019decoding, tam2019human} for reviews of models for decoding movement). Simpler approaches based on decoding of discrete classes use various classification algorithms, such as linear discriminant analysis, k-nearest neighbors, support vector machines and others. Models that decode continuous kinematic trajectories can use linear methods, such as the linear regression. For decoding of position and velocity of movement, models of linear dynamical systems, such as the Kalman filter, are typically employed~\cite{pistohl2008prediction, kellis2012decoding}. They compute hidden states of the system, such as intended movement, based on movement from the previous time stamp, and model observed states, such as neural activity, as linearly dependent on the computed hidden state data. Noise associated with observed and hidden states is modeled separately. Such formulation allows for better estimation of time series data and helps boost performance on noisy non-stationary iEEG data. Next to the Kalman filter, hidden Markov models are also often used to handle more complex temporal dynamics in a model with hidden and observable states~\cite{benabid2019exoskeleton}. For better temporal stability adaptive and switching decoding models~\cite{benabid2019exoskeleton} have been employed as well.

Recently, computational approaches based on artificial neural network approaches~\cite{schwemmer2018meeting, xie2018decoding, wang2018ajile, pan2018rapid, skomrock2018characterization, choi2021non, sliwowski2022decoding} have become increasingly popular. These models consistently appear to provide more accurate and robust decoding compared to simpler baselines~\cite{schwemmer2018meeting, xie2018decoding, sliwowski2022decoding}. Moreover, these approaches have the potential to handle concerns associated with real-time decoding for practical BCI applications. Namely, they can provide superior speed of the BCI response and stable performance over sessions and days~\cite{schwemmer2018meeting}, while offering relatively computationally inexpensive, and even power-efficient solutions. The latter is further explored by development of spiking neural networks in combination with power-efficient memristive hardware as BCI solutions~\cite{liu2020neural}. Another interesting AI application in motor decoding is analysis of video recordings collected simultaneously with ECoG and data-driven extraction of movement feature from video data~\cite{wang2018ajile} for decoding from the brain. Thus, similar to communication, external AI tools for extracting relevant features and building physical models can guide motor decoding and further improve the state-of-the-art performance. Altogether, as in the case of BCI for communication, it appears that state-of-the-art BCIs for motor control, implemented either on microelectrode or iEEG implant, will continue to be powered by AI decoding algorithms and external tools for achieving highly accurate, sophisticated, reliable control of external devices by paralyzed individuals.


\subsection{iEEG for deep brain stimulation}
\label{subsec:dbs}

Invasive neurotechnology for healthcare does not only focus on restoration of a lost
function, such as speech, hearing, vision or motor control. Another branch of invasive neurotechnology targets chronic neurological and psychiatric conditions that do not respond to conventional treatment. Such conditions include movement disorders, such as Parkinson's disease, essential tremor, dystonia (involuntary muscle contractions); epilepsy; mental disorders, such as major depression, schizophrenia, obsessive-compulsive disorder, chronic pain and others. These conditions affect large numbers of people. For example, it is estimated that over 160 million people worldwide suffer from major depression~\cite{james2018global}. About 50-80 million worldwide suffer from epilepsy~\cite{brodie1997commission,kwan2000early}, of which 30-40\% do not respond to medication~\cite{kobau2008epilepsy}). Parkinson's disease affects over 6 million of the general population~\cite{vos2016global}.

Many of these conditions, as well as the affected behavioral and cognitive function, have been associated with neural activity in deeper brain structures: basal ganglia in Parkinson's disease~\cite{graybiel2000basal, obeso2008functional}, nucleus accumbens and median forebrain bundle in major depression~\cite{schlaepfer2014deep}, amygdala and hippocampus in temporal lobe epilepsy~\cite{laxpati2014deep}. Severe cases that are resistant to medication can be treated with an invasive form of therapy called \emph{deep brain stimulation} (DBS). This neurotechnology assumes implantation of a small electrode wire connected to a pulse generation device, typically placed under the skin around the chest~\cite{coffey2009deep}, to modulate function of deeper brain structures. The device attempts to inhibit pathological neural activity via electrical stimulation of the neural tissue surrounding the implant and thereby improve the patient's condition.

Among the earliest applications of an implantable DBS technology in patients with Parkinson's disease was work by Benabid and colleagues~\cite{benabid1987combined}. The authors showed that high-frequency stimulation of ventral intermediate nucleus resulted in up to 88\% improvement of tremor symptoms. The effect persisted up to 29 months after implantation. Subsequently, DBS in subthalamic nuclei was developed and successfully tested in patients with Parkinson's disease~\cite{pollak1993effects, limousin1995effect}. It has since been shown to provide more relief and emotional well-being to patients compared to conventional medical treatment~\cite{deuschl2006randomized, williams2010deep, odekerken2013subthalamic}, and is now considered one of the state-of-the-art solutions for treating Parkinson's disease. Analogous applications were developed and successfully demonstrated in patients with medication-resistant epilepsy. Earlier work reported that modulation of anterior thalamic nucleus could lead to reduction in seizure frequency~\cite{cooper1980reversibility, andrade2006long}. Following successful large-scale trials by Fisher and colleagues~\cite{fisher2010electrical}, most modern DBS devices target this region for reduction of epileptic seizures. At the same time, stimulation of other deep brain structures, such as hippocampal formation~\cite{boex2011chronic}, subthalamic nucleus and substantia nigra~\cite{handforth2006deep}, centromedian nucleus~\cite{valentin2013deep} and cerebellar regions~\cite{velasco2005double} has also been explored. See~\cite{loddenkemper2001deep, theodore2004brain, laxpati2014deep} for reviews. Apart from Parkinson's disease and epilepsy, DBS applications for tackling symptoms of chronic mental disorders are also being developed. These include DBS treatment of major depression~\cite{mayberg2005deep}, obsessive-compulsive disorder~\cite{blomstedt2013deep} and other conditions~\cite{perlmutter2006deep,lozano2019deep}.

DBS systems have already been able to help many people worldwide. Yet despite the accumulated success, commercially available DBS implementations have several limitations. Most systems are based on a single wire with a few electrodes (typically four)~\cite{coffey2009deep} and therefore provide only limited brain coverage, that may not be suitable for comprehensive monitoring of the pathology, as brain signals in health and disease are believed to reflect large-scale states of dynamic neural systems~\cite{richardson2012large, breakspear2017dynamic}. Moreover, commercially available DBS systems are not suitable for using recorded signals (with the exception of the Percept PC~\cite{goyal2021development} and the Activa PC+S (Medtronic) neurostimulators), in guiding the stimulation strategy (closed-loop solutions). Instead, most systems rely on open-loop solutions, which means that they deliver a constant stimulation pulse to the implanted tissue based on a simple trial and error approach~\cite{kringelbach2007translational, lozano2019deep} and do not process observed brain signals to adapt their stimulation behavior~\cite{kringelbach2007translational, sun2014closed}. Such DBS systems are less effective and could drain the battery during continuous stimulation even when it may not be necessary. As a consequence, DBS shows high variability of success and can lead to DBS-induced adverse effects~\cite{kupsch2006pallidal, zauber2009stimulation, follett2010pallidal, schrader2011gpi}. Altogether, there is a need for better understanding of the neural signal, larger-scale recordings and active, adaptive monitoring of the system. Next-generation devices could represent a closed-loop system based on neuromodulation, which means that they could continuously monitor brain activity in compromised regions, identify neural patterns associated with healthy and pathological states, and devise a stimulation plan for treatment (see also Chapters 48, 59 and 61).  

A number of prototypes for adaptive closed-loop solutions have already been demonstrated~\cite{yamamoto2013demand, malekmohammadi2016kinematic, herron2016chronic, swann2018adaptive}. Among those are studies that employ iEEG recordings for continuous recording and processing of neural signals to observe pathological activity and effects of stimulation, and adapt stimulation parameters accordingly. For example, several recent studies used ECoG grids implanted in addition to DBS wires for monitoring movement-related activity on the motor cortex~\cite{herron2016chronic, swann2018adaptive}. Herron et al. used ECoG electrodes to detect intended hand movements in a patient with essential tremor~\cite{herron2016chronic}. In this disorder, tremor occurs primarily during intentional movement, and applying stimulation after movement intention has been identified has led to a more energy-efficient DBS system without losing much of the therapeutic value. Similar results were obtained in patients with Parkinson's disease by Swann et al., who monitored high gamma ECoG activity on the motor cortex for markers of involuntary movements  called dyskinesia, and used them to update DBS parameters. Another ECoG-based monitoring system for adaptive stimulation was tested in epilepsy and showed an up to 70\% reduction of seizure frequency and persisting long-term positive effects~\cite{jobst2017brain}.  

In the context of DBS, another type of iEEG recordings is starting be be particularly attractive. As a less invasive technology compared to ECoG, designed for monitoring neural activity in deeper brain regions, sEEG has a lot of potential for successful guidance of adaptive DBS systems. A combination of continuous registration of activity throughout the brain and advanced computational algorithms for processing large-scale multi-channel data with complex temporal dynamics and connectivity structure, may provide the necessary understanding of the nature of neural signals in health and disease and ultimately provide appropriate DBS treatment. For example, in a recent work, Scangos et al. implanted sEEG electrodes in an individual with major depression to monitor neural signals in their hippocampus, amygdala, orbitofrontal, cingulate and striatum cortex~\cite{scangos2021closed}. The researchers were able to train online classifiers to continuously monitor sEEG activity in these regions and identify patient-specific biomarkers of pathological states. This allowed them to link neural activity in the patient's amygdala to their most severe depression symptoms. This knowledge was used to guide the subsequently implanted DBS system by detecting previously identified biomarkers of depression and applying stimulation to the affected sites 
. The study showed that this approach resulted in a significant improvement of the patient's symptoms. Similarly, in another study, the authors demonstrated a successful mapping of neural states monitored with sEEG to individual symptoms of the underlying treatment-resistant depression~\cite{sheth2021deep}. Moreover, the authors used sEEG data to optimize parameters of stimulation applied to the affected brain regions via a standard DBS implant. SEEG-informed stimulation led to a large improvement of the patient's symptoms and their overall quality of life. Altogether, this work demonstrates the potential of sEEG implants for development of the next-generation closed-loop DBS systems with data-driven neuromodulation and personalized approaches to real-world clinical applications. 


A line of research that takes this work one step further is based on the development of iEEG-based DBS systems, an integrated bidirectional approach to brain monitoring and stimulation through the same set of iEEG electrodes. Two recent studies have demonstrated a proof of concept for for such a device. Basu et al. developed a system for monitoring cognitive control in individuals with mental disorders based on decoding of neural events associated with behavioral task performance~\cite{basu2021closed}. Saucedo et al. reported on a similar approach applied to epilepsy and hippocampal sclerosis~\cite{saucedo2022optimizing}.


An important component of these adaptive DBS systems is machine learning and AI models. It has been shown that advanced algorithms that can handle high dimensionality of large-scale recordings and temporal connectivity structure may be better suited for signal processing and identification of brain states in health and disease. These algorithms are being increasingly used for identification of biomarkers of healthy and pathological states in real time and subsequent adaptation of stimulation parameters depending on how they affect behavior~\cite{neumann2021machine, merk2022electrocorticography}. In several applications, the potential of AI models for making complex predictions can be particularly valuable, for example, for prediction of seizure onset in epilepsy~\cite{kuhlmann2018seizure} or symptom severity in major depression~\cite{scangos2021closed}. Recently, progress in this area has been accelerated by an increase in availability of large public datasets during clinical recordings of pathological and healthy brain activity. The latter has stimulated the emergence of data science competitions, where data analysts compete in developing the algorithm that best solves the task, for example, seizure prediction. Next to that, the theory and computational models of large-scale network activity in health and disease are growing ever more comprehensive and complex~\cite{sussillo2016lfads, breakspear2017dynamic, meier2022virtual}. These sophisticated models of dynamical systems and neural connectomics have the potential to drive forward our understanding of complex neural processes monitored with continuous multi-channel recordings, such as iEEG. In DBS neurotechnology, these contributions can lead to novel closed-loop real-world solutions that provide more effective personalized treatments for severe chronic neurological conditions and thereby substantially improve quality of life.  



\section*{Acknowledgements}

This work was supported by the Netherlands Organisation for Scientific Research (NWO) and is part of the Language in Interaction consortium (NWO Gravitation Grant No. 024.001.006) and INTENSE consortium (NWO grant No. 17619). We thank Mariska Vansteensel, Jordy Thielen, Linda Geerligs and Pieter Kubben for their helpful comments on the initial version of the manuscript.

\bibliography{references.bib}
\bibliographystyle{unsrt}

\end{document}